\documentclass[preprint,12pt]{elsarticle}

\usepackage{amssymb}
\usepackage{amsmath}
\usepackage{hyperref}

\usepackage[percent]{overpic}

\journal{Chaos, Solitons \& Fractals}

\begin{document}

\begin{frontmatter}



\title{Heterogeneity-induced chaos in globally coupled Stuart--Landau oscillators}


\author{Luekai Zhao}

\author{Nariya Uchida}%
\ead{nariya.uchida@tohoku.ac.jp}

\affiliation{organization={Department of Physics, Tohoku University},
            city={Sendai},
            postcode={980-8578}, 
            country={Japan}}


\begin{abstract}
We show that linear-growth-rate heterogeneity combined with complex global coupling can generate quasiperiodicity and chaos in a finite population of Stuart--Landau oscillators with identical natural frequencies and no Kerr-type nonlinear frequency shift. 
Lyapunov spectra identify limit cycles, quasiperiodic tori, and chaotic attractors, while the order-parameter amplitude exhibits signatures consistent with period-doubling-like transitions towards chaos. 
We further derive a finite-size invariant manifold of vanishing-order-parameter phase-locked solutions represented by closed polygons with fixed side lengths and demonstrate their configuration-dependent transverse stability. 
Ensemble sampling reveals pronounced coexistence among polygonal and ordinary phase-locked cycles, quasiperiodic tori, and chaotic attractors. 
An amplitude--time rescaling identifies the leading parameter approximately organizing the regime boundaries and the similar bifurcation and multistability patterns. 
These results advance our understanding of collective dynamics in amplitude-inclusive oscillator networks by revealing how heterogeneity and complex coupling jointly shape dynamical complexity, bifurcation structure, and multistability.
\end{abstract}




\end{frontmatter}


\section{Introduction}
\label{Intro}

The Stuart--Landau (SL) oscillator~\cite{landau1944problem,stuart1960non} is the normal form of a self-sustained oscillator near a Hopf bifurcation and therefore provides a minimal amplitude--phase description of limit-cycle dynamics~\cite{kuramoto1984chemical}, thus making it suitable for studying collective phenomena such as synchronization~\cite{pikovsky2001synchronization,kuete2026synchronization}, amplitude death~\cite{zhao2018enhancing,zou2014emergence} and clustering~\cite{kemeth2019cluster,kemeth20212} in oscillator networks.
Beyond their role in synchronization theory, SL oscillators are applied to a broad range of contexts including biology and optical and electrical experiments~\cite{daido2004aging,klar2023scale,hatakeyama2015reciprocity,harrison2020synchronization,gambuzza2020experimental}. 
In particular, SL oscillator networks have been used as reduced models of large-scale brain dynamics~\cite{deco2017dynamics,deco2015rethinking}. 

In addition to synchronization, complex dynamics also emerge in SL oscillators and are often associated with Kerr-type nonlinearity, or equivalently a nonlinear frequency shift~\cite{shing2013phase}. 
In this case, the instantaneous angular velocity depends on the oscillation amplitude, so amplitude fluctuations can be converted into phase modulations. 
This amplitude--phase conversion is a well-known route to complex collective behavior, including quasiperiodicity and chaos~\cite{ku2015dynamical,pandey2026complex,premalatha2018stable,hohlein2019lyapunov}. 
Recent work has shown that nonreciprocal coupling can destabilize synchronized periodic motion~\cite{ryu2026nonreciprocity}; however, chaos emerges in that model only when a nonlinear frequency shift is present.
This suggests a key question: is a nonlinear frequency shift necessary for chaos generation in coupled SL oscillators?

Heterogeneity offers a natural alternative route, as it can reorganize collective oscillator dynamics without changing the basic limit-cycle nature of the individual units~\cite{shing2013phase}.
In the absence of a nonlinear frequency shift, heterogeneity can enter mainly through either natural frequencies or linear growth rates.
Natural-frequency heterogeneity has been extensively studied: Shiino et al.~\cite{shiino1989synchronization} derived a phase diagram including synchronization, amplitude death, incoherence and desynchronization, whereas Matthews et al.~\cite{matthews1990phase} further identified the desynchronization as complex behavior including quasiperiodicity and chaos.
By contrast, previous studies of linear-growth-rate heterogeneity have mainly focused on synchronization~\cite{millan2025synchronization} and aging transitions~\cite{daido2004aging}.
It therefore remains unclear whether linear-growth-rate heterogeneity can generate chaos in coupled SL oscillators without dependence on a nonlinear frequency shift.

In this work, we address this question by showing that linear-growth-rate heterogeneity and complex global coupling are sufficient to generate chaos in a finite SL oscillator system even when all natural frequencies are identical and there is no nonlinear frequency shift.
We construct the dynamical-regime diagram and identify the underlying mechanism in which heterogeneity and coupling jointly destabilize phase locking through a mismatch-driven feedback. 
Exploiting global rotational symmetry, we reveal a unified bifurcation structure connecting phase-locked motion and chaos through closely related transition routes. 
Beyond the regime map, we uncover a family of vanishing-order-parameter polygonal states whose existence and stability enrich the set of possible collective organizations beyond conventional synchronized dynamics. 
Furthermore, ensemble sampling reveals the coexistence of phase-locked cycles, polygonal states, quasiperiodic tori, and chaotic attractors, while the approximate scaling identifies a leading parameter that organizes the regime boundaries, bifurcation structure, and multistability. 
Taken together, these results provide a comprehensive dynamical picture of the system by linking the bifurcation to chaos, attractor coexistence, and the approximate scaling structure of the parameter space.

\section{Model}
\label{sec:model}

We consider a system of $M$ globally coupled SL oscillators.
The state of the $j$th oscillator is represented by the complex variable
$z_j=x_j+\mathrm{i}y_j$, whose dynamics is governed by
\begin{equation}
\dot{z}_j
=
\left[
\beta_j+\mathrm{i}\omega
-
\left(k_r+\mathrm{i}k_i\right)|z_j|^2
\right]z_j
+
\left(C_r+\mathrm{i}C_i\right)
\sum_{\substack{\ell=1\\ \ell\neq j}}^{M}
\left(z_\ell-z_j\right),
\qquad
j=1,\ldots,M.
\label{eq:model}
\end{equation}
Here, $\beta_j$ denotes the linear growth rate of the $j$th oscillator,
$\omega$ is the common natural angular frequency. 
$k_r>0$ causes dissipative nonlinearities, determining the nonlinear saturation of the oscillation amplitude. 
$k_i$ brings about Kerr-type nonlinearities that quantify the dependence of the oscillation frequency on the amplitude. 
The real and imaginary parts of the complex coupling coefficient, $C_r$ and $C_i$, respectively describe the dissipative and reactive components of the global coupling.

The heterogeneity is introduced through a linear variation of the growth
rates,
\begin{equation}
\beta_j
=
\beta_{\min}
+
\frac{j-1}{M-1}\Delta\beta.
\label{eq:beta_distribution}
\end{equation}
where $\Delta\beta=\beta_{\max}-\beta_{\min}$.

The identical-oscillator limit is recovered for $\Delta\beta=0$, whereas
a finite $\Delta\beta$ introduces amplitude heterogeneity into the
uncoupled oscillators.

For an isolated oscillator with $\beta_j>0$, the stable limit-cycle
amplitude and angular frequency are given by
\begin{equation}
r_j^{\mathrm{uncoupled}}
=
\sqrt{\frac{\beta_j}{k_r}},
\qquad
\Omega_j^{\mathrm{uncoupled}}
=
\omega-\frac{k_i}{k_r}\beta_j.
\label{eq:isolated_solution}
\end{equation}
Thus, when $k_i=0$, all isolated oscillators have the same angular frequency despite having different limit-cycle amplitudes. 
In this case, any frequency differentiation or chaotic dynamics observed in the
coupled system originates from the collective interaction between heterogeneity and complex global coupling (dissipative and reactive), rather than from Kerr-type nonlinearities or an imposed distribution of natural frequencies.

To characterize the collective dynamics, we introduce the complex order parameter
\begin{equation}
Z(t)
=
\frac{1}{M}\sum_{j=1}^{M}z_j(t),
\label{eq:order_parameter}
\end{equation}
and Eq.~\eqref{eq:model} becomes
\begin{equation}
\dot{z}_j
=
\left[
\beta_j+\mathrm{i}\omega
-
\left(k_r+\mathrm{i}k_i\right)|z_j|^2
\right]z_j
+
M\left(C_r+\mathrm{i}C_i\right)
\left(Z-z_j\right).
\label{eq:model_mean_field}
\end{equation}

The equations are numerically integrated using the fourth-order Runge--Kutta method with a time step $\Delta t=0.01$. 
The parameters are fixed at $M=8$, $\beta_{\min}=0.01$, $\omega=0.2$, $k_r=0.01$, and $k_i=0$.
A transient of at least 20000 time units is discarded before calculating Lyapunov spectrum~\cite{benettin1980lyapunov,sandri1996numerical}, frequency spectra, and Lyapunov distribution.
For initial conditions, the real and imaginary parts of $z_j(0)$ were independently drawn from the uniform distribution $\mathcal{U}[-5,5]$ through random seeds. 
For all single-initial-condition calculations, the same fixed initial condition was used at every parameter point; hereafter, it is referred to as the prescribed initial condition.
The corresponding random seed and details of the numerical implementation are provided in the accompanying codes.

\section{Dynamical regimes in parameter space}
\label{sec:regimes}

We first clarify why a reactive component of the coupling is essential
for the emergence of complex dynamics in the present model.
Consider the purely real-coupling reference system with $C_i=0$. In the frame rotating with the common natural frequency
$\omega$,
\begin{equation}
w_j(t)=z_j(t)e^{-\mathrm{i}\omega t},
\end{equation}
the equations reduce to
\begin{equation}
\dot{w}_j
=
\left(\beta_j-k_r|w_j|^2\right)w_j
+
C_r\sum_{\ell\ne j}(w_\ell-w_j).
\label{eq:real_coupling_rotating_frame}
\end{equation}
Writing $w_j=x_j+\mathrm{i}y_j$, this system is a gradient flow,
\begin{equation}
\dot{\boldsymbol q}
=
-\nabla\mathcal{V}(\boldsymbol q),
\qquad
\boldsymbol q=(x_1,y_1,\ldots,x_M,y_M)^{\mathsf T},
\end{equation}
with the potential
\begin{equation}
\mathcal{V}
=
\sum_{j=1}^{M}
\left[
-\frac{\beta_j}{2}|w_j|^2
+
\frac{k_r}{4}|w_j|^4
\right]
+
\frac{C_r}{2}
\sum_{j<\ell}|w_j-w_\ell|^2 .
\label{eq:real_coupling_potential}
\end{equation}
Indeed,
\begin{equation}
-\frac{\partial\mathcal{V}}{\partial x_j}
=
\left[\beta_j-k_r(x_j^2+y_j^2)\right]x_j
+
C_r\sum_{\ell\ne j}(x_\ell-x_j),
\end{equation}
and the same relation holds for $y_j$. Therefore,
\begin{equation}
\frac{\mathrm{d}\mathcal{V}}{\mathrm{d}t}
=
-\sum_{j=1}^{M}|\dot{w}_j|^2
\leq0.
\label{eq:potential_decay}
\end{equation}
Thus, the asymptotic dynamics of the purely real-coupled system is restricted to equilibria in the rotating frame. 
In the original frame, these equilibria correspond to fixed points or limit cycles.
Nontrivial quasiperiodic tori and sustained chaotic attractors are therefore excluded in this case. 
A nonzero reactive coupling $C_i$ breaks this gradient-flow structure and enhances feedback between amplitude and phase degrees of freedom.

We now characterize the dynamics in the parameter plane spanned by the reactive coupling strength $C_i$ and the heterogeneity $\Delta\beta$. 
Throughout this work, the dissipative and reactive
coupling components are varied at the fixed ratio
\begin{equation}
C_r=0.25C_i.
\label{eq:coupling_ratio}
\end{equation}
Thus, varying $C_i$ changes the magnitude of the complex coupling coefficient while preserving its argument. 

At each parameter point, the attractor reached from the prescribed initial
condition is classified using the four largest Lyapunov exponents,
\begin{equation}
\lambda_1\geq\lambda_2\geq\lambda_3\geq\lambda_4.
\end{equation}
\begin{figure}[t]
    \centering
    \includegraphics[width=0.98\linewidth]
    {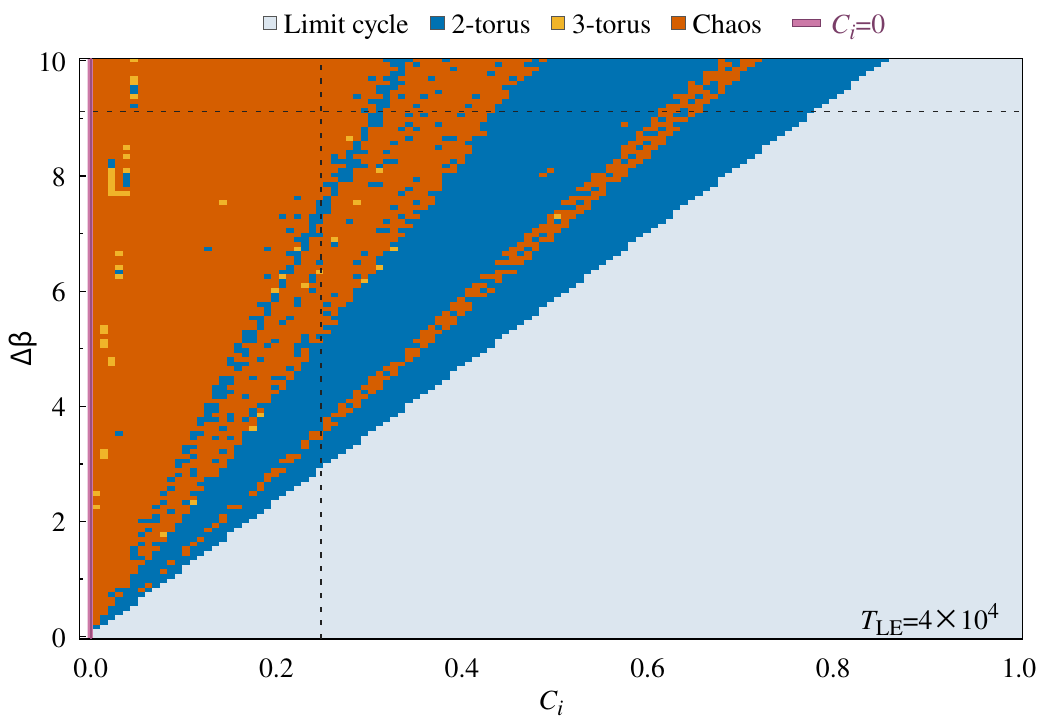}
    \caption{
    Lyapunov-based dynamical-regime map in the
    $(C_i,\Delta\beta)$ parameter plane. The observed attractors are
    classified as phase-locked limit cycles (light gray),
    two-dimensional tori (blue), three-dimensional tori (yellow), and
    chaotic attractors (orange). The magenta line at $C_i=0$ represents
    the uncoupled product of limit cycles, where $C_r=C_i=0$ under
    Eq.~\eqref{eq:coupling_ratio}. The vertical and horizontal dashed
    lines indicate the cuts $C_i=0.248$ and $\Delta\beta=9.12$,
    respectively. The classification at each point refers to the
    attractor selected by the prescribed initial condition and therefore
    does not exclude multistability. The post-transient calculation lasts for $T_{\text {LE}}=4\times10^4$ time units. 
    }
    \label{fig:bifurcation_landscape}
\end{figure}
A Lyapunov exponent is regarded as numerically zero when
\begin{equation}
|\lambda_j|\leq\varepsilon_{\mathrm{LE}},
\qquad
\varepsilon_{\mathrm{LE}}=3\times10^{-4}.
\end{equation}
Details of the numerical tolerance $\varepsilon_{\mathrm{LE}}$ and convergence checks for the finite-time Lyapunov exponents are provided in~\ref{app:lle_convergence}.
The numerical classification is
\begin{align}
\text{limit cycle:}\qquad&
|\lambda_1|\leq\varepsilon_{\mathrm{LE}},
\quad
\lambda_2<-\varepsilon_{\mathrm{LE}},
\nonumber\\
\text{two-torus:}\qquad&
|\lambda_1|,|\lambda_2|
\leq\varepsilon_{\mathrm{LE}},
\quad
\lambda_3<-\varepsilon_{\mathrm{LE}},
\nonumber\\
\text{three-torus:}\qquad&
|\lambda_1|,|\lambda_2|,|\lambda_3|
\leq\varepsilon_{\mathrm{LE}},
\quad
\lambda_4<-\varepsilon_{\mathrm{LE}},
\nonumber\\
\text{chaos:}\qquad&
\lambda_1>\varepsilon_{\mathrm{LE}}.
\label{eq:LE_classification}
\end{align}
Here, only the four largest Lyapunov exponents are used because all remaining exponents are negative for every attractor reached from the prescribed initial condition. 
A transversely stable polygonal manifold would be classified separately according to its number of neutral directions:
\begin{equation}
\text{\(n\)-dimensional polygonal manifold:}\qquad
\max_{1\leq j\leq n}|\lambda_j|
\leq\varepsilon_{\mathrm{LE}},
\quad
\lambda_{n+1}<-\varepsilon_{\mathrm{LE}}.
\end{equation}
As demonstrated in Sec.~\ref{sec:polygonal_manifold}, however, none of the attractors selected by the prescribed initial condition lies on the polygonal manifold.

Figure~\ref{fig:bifurcation_landscape} shows that heterogeneity and complex global coupling generate several distinct dynamical regimes.
Increasing $\Delta\beta$ at fixed coupling produces extended quasiperiodic and chaotic regions, whereas increasing the coupling strength at fixed heterogeneity generally drives the system back toward the limit-cycle regime.
Several studies have likewise reported a transition toward phase-locked states with increasing coupling strength~\cite{shing2013phase,millan2025synchronization,matthews1990phase,shiino1989synchronization}, suggesting that this tendency is a well-established feature of coupled SL systems.
Three-dimensional tori appear only in comparatively narrow regions, mainly near the interface between two-torus and chaotic dynamics.

The mechanism can be understood from the mean-field form of the
equations. Introducing
\begin{equation}
z_j=r_j e^{\mathrm{i}\theta_j},
\qquad
Z=R e^{\mathrm{i}\Theta},
\qquad
\psi_j=\Theta-\theta_j,
\end{equation}
Eq.~\eqref{eq:model_mean_field} gives
\begin{align}
\dot{r}_j
={}&
\left(\beta_j-k_r r_j^2\right)r_j
+
M C_r\left(R\cos\psi_j-r_j\right)
-
M C_i R\sin\psi_j,
\label{eq:radial_mean_field}\\
\dot{\theta}_j
={}&
\omega
+
M C_r\frac{R}{r_j}\sin\psi_j
+
M C_i
\left(
\frac{R}{r_j}\cos\psi_j-1
\right).
\label{eq:phase_mean_field}
\end{align}
Without coupling, the $j$th oscillator is attracted to the preferred amplitude $r_j^{\mathrm{uncoupled}}$, Eq.~\eqref{eq:isolated_solution}.
Thus, increasing $\Delta\beta$ broadens the distribution of preferred amplitudes and enhances the mismatch between individual oscillators and the collective mean field.

\begin{figure}[!htbp]
    \centering
    \includegraphics[width=1.0\linewidth]
    {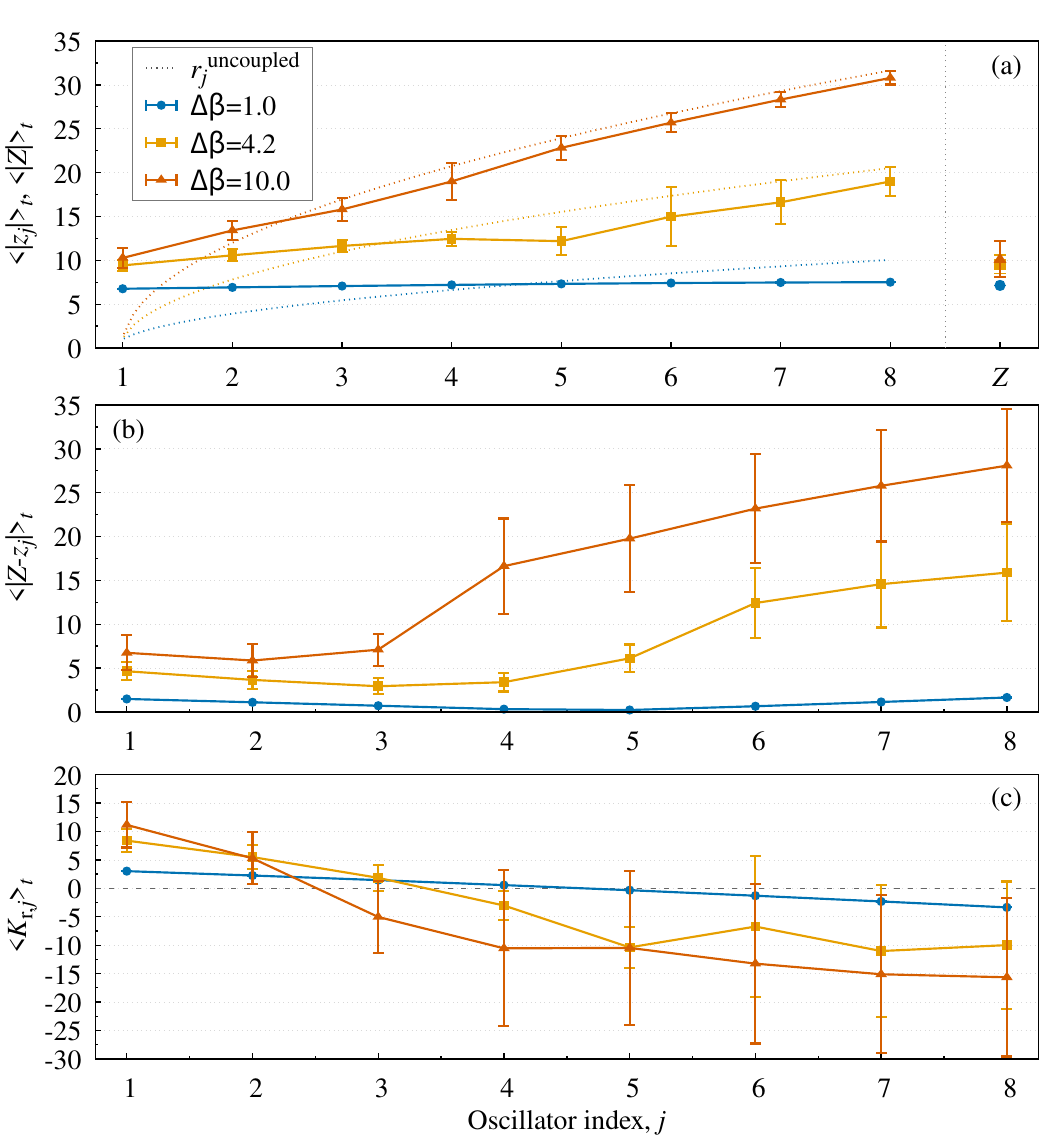}
    \caption{
    Mechanistic quantities along the cut $C_i=0.248$ for
    $\Delta\beta=1.0,4.2,10.0$.
    (a) Time-averaged oscillator amplitudes
    $\langle |z_j| \rangle_t$, order-parameter amplitude
    $\langle |Z| \rangle_t$, and uncoupled amplitudes
    $r_j^{\mathrm{uncoupled}}$.
    (b) Mean-field mismatch
    $\langle |Z-z_j| \rangle_t$, which measures the displacement entering
    the global diffusive coupling term.
    (c) Signed radial projection
    $\langle K_{r,j}\rangle_t$ of the complex coupling term. Positive
    and negative values correspond to radial injection and suppression,
    respectively.
    In all three panels, error bars denote temporal standard deviations along the
    post-transient trajectory.
    }
    \label{fig:mechanism_triplet}
\end{figure}

This mechanism is quantified along the vertical cut $C_i=0.248$ using three representative heterogeneity strengths,
\begin{equation}
\Delta\beta=1.0,\quad 4.2,\quad 10.0.
\end{equation}
For each case, we measure time--averaged quantities
$\langle |z_j| \rangle_t$, $\langle |Z| \rangle_t$,
$\langle |Z-z_j| \rangle_t$, and
$\langle K_{r,j} \rangle_t$, where
\begin{equation}
K_{r,j}(t)
=
\operatorname{Re}
\left[
(C_r+\mathrm{i}C_i)M(Z-z_j)
\frac{z_j^*}{|z_j|}
\right]
\label{eq:radial_projection}
\end{equation}
is the signed radial projection of the complex coupling force.

Figure~\ref{fig:mechanism_triplet} shows the basic trend. 
Increasing $\Delta\beta$ broadens the local amplitude profile and enlarges the mean-field mismatch $\langle |Z-z_j| \rangle_t$. 
The complex coupling coefficient $C_r+\mathrm{i}C_i$ rotates this mismatch in the complex
plane, thereby converting it into an enhanced magnitude of the time-averaged radial coupling force
$\left|\langle K_{r,j}\rangle_t\right|$.

This gives the first part of the mechanism:
\begin{equation}
\Delta\beta \uparrow
\quad\Rightarrow\quad
\langle |Z-z_j| \rangle_t \uparrow
\quad
\xrightarrow{\;C_r+\mathrm{i}C_i\;}
\quad
\left|\langle K_{r,j}\rangle_t\right|\uparrow .
\end{equation}

The resulting dynamics is determined by the competition between this radial coupling force and the local radial restoring force
\begin{equation}
F_j(r_j)
=
\left(\beta_j-k_r r_j^2\right)r_j .
\end{equation}
Expanding around $r_j^{\mathrm{uncoupled}}$ gives
\begin{equation}
F_j\left(r_j^{\mathrm{uncoupled}}+\delta r_j\right)
=
-2\beta_j\delta r_j
-3k_r\sqrt{\frac{\beta_j}{k_r}}\,\delta r_j^2
-k_r\delta r_j^3 .
\label{eq:radial_taylor}
\end{equation}
Thus, the local radial stiffness at $r_j^{\mathrm{uncoupled}}$ is
\begin{equation}
-F'_j\left(r_j^{\mathrm{uncoupled}}\right)=2\beta_j.
\end{equation}
Accordingly, a larger $\beta_j$ yields a greater local radial stiffness, while increasing $\Delta\beta$ makes the restoring response increasingly heterogeneous across oscillators.
Heterogeneity therefore enhances both sides of the radial competition: the mean-field radial coupling through $|Z-z_j|$ and the local radial restoring stiffness through $\beta_j$.

A phase-locked limit cycle corresponds to a static self-consistent balance between $K_{r,j}$ and $F_j$. 
This requires not only
\begin{equation}
F_j(r_j^*)+K_{r,j}^*=0
\qquad
\text{for all }j,
\label{static balance radial}
\end{equation}
but also a common angular velocity,
\begin{equation}
\dot{\theta}_1=\dot{\theta}_2=\cdots=\dot{\theta}_M=\Omega .
\label{static balance angular}
\end{equation}
These two conditions are restrictive: the radial forces must balance for all oscillators while the angular dynamics simultaneously supports fixed relative phases. 
When they are satisfied, the system settles to a shifted stationary amplitude profile, giving a phase-locked limit cycle.

As $\Delta\beta$ increases, both the mismatch-driven radial coupling $K_{r,j}$ and the heterogeneous radial restoration $F_j$ become stronger in magnitude.
These radial effects are coupled to the angular dynamics because $K_{r,j}$ depends on the phase difference $\psi_j$, whereas $\dot{\theta}_j$ in Eq.~\eqref{eq:phase_mean_field} depends on the radius $r_j$. 
A radial perturbation therefore changes the angular motion, and the resulting phase change feeds back into the radial projection. 
When this radial--angular feedback remains stabilized, the system forms a phase-locked limit cycle. 
When the static balance loses stability, the balance becomes time dependent, leading to quasiperiodic or chaotic collective motion.

Fig.~\ref{fig:bifurcation_landscape} shows the dynamical regime selected by one prescribed initial condition, not the complete set of coexisting attractors. 
This limitation motivates the analyses that follow. 
We first resolve the bifurcation routes along the two dashed cuts using a well-defined observable, then identify a family of solutions with vanishing order parameter that is absent in Fig.~\ref{fig:bifurcation_landscape}, and finally employ an ensemble of initial conditions to expose the multistability concealed by the single-initial-condition regime map.

\section{Rotational symmetry and bifurcation to chaos}
\label{sec:rotational_symmetry}

The globally coupled SL system is equivariant under a uniform rotation of all oscillator states. 
Denoting the full state by $\mathbf{z}=(z_1,\ldots,z_M)^{\mathsf T}$, the rotational transformation is defined as
\begin{equation}
g_{\psi}\mathbf{z}
=
e^{\mathrm{i}\psi}\mathbf{z},
\qquad
\psi\in[0,2\pi).
\end{equation}
The vector field $\mathbf{F}$ associated with Eq.~\eqref{eq:model} satisfies
\begin{equation}
\mathbf{F}(g_{\psi}\mathbf{z})
=
g_{\psi}\mathbf{F}(\mathbf{z}),
\label{eq:rotational_equivariance}
\end{equation}
because $|e^{\mathrm{i}\psi}z_j|=|z_j|$, while every onsite and coupling term acquires the same phase factor $e^{\mathrm{i}\psi}$. 
Thus, if $\mathbf{z}(t)$ is a solution, then
$e^{\mathrm{i}\psi}\mathbf{z}(t)$ is also a solution.

For any nonzero state, the continuous set
\begin{equation}
\mathcal{O}(\mathbf{z})
=
\left\{
e^{\mathrm{i}\psi}\mathbf{z}
\mid
0\leq\psi<2\pi
\right\}
\end{equation}
forms a group orbit topologically equivalent to a circle. 
Consequently, an isolated, connected, one-dimensional attractor invariant
under the rotational symmetry must be such a circular group orbit. 
A phase-locked limit cycle belongs to this class: the oscillator amplitudes and relative phases remain constant, while the complete configuration rotates uniformly.

\begin{figure}[!htbp]
    \centering

    \begin{minipage}[t]{0.498\linewidth}
        \centering
        \includegraphics[width=\linewidth]{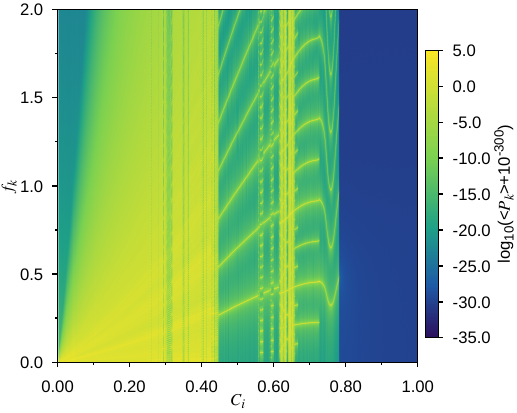}
        \par\vspace{-0.3em}
        {\scriptsize\textbf{(a)}}
    \end{minipage}%
    \hfill
    \begin{minipage}[t]{0.498\linewidth}
        \centering
        \includegraphics[width=\linewidth]{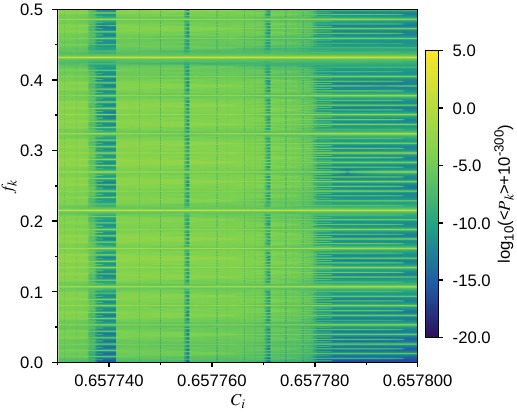}
        \par\vspace{-0.3em}
        {\scriptsize\textbf{(b)}}
    \end{minipage}

    \par\vspace{0.2em}

    \begin{minipage}[t]{0.498\linewidth}
        \centering
        \includegraphics[width=\linewidth]
        {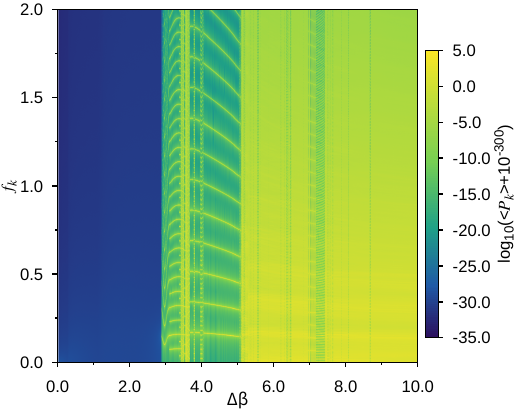}
        \par\vspace{-0.3em}
        {\scriptsize\textbf{(c)}}
    \end{minipage}%
    \hfill
    \begin{minipage}[t]{0.498\linewidth}
        \centering
        \includegraphics[width=\linewidth]
        {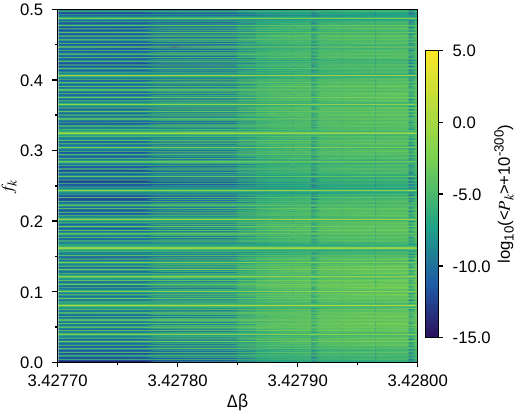}
        \par\vspace{-0.3em}
        {\scriptsize\textbf{(d)}}
    \end{minipage}

    \caption{
    Power spectral density of the order-parameter modulus
    $R(t)=|Z(t)|$ along the two parameter cuts marked in
    Fig.~\ref{fig:bifurcation_landscape}. The color denotes
    $\log_{10}(\langle {P}_k \rangle+10^{-300})$.
    (a) Spectrum as a function of $C_i$ at
    $\Delta\beta=9.12$.
    (b) Enlarged view of a narrow interval $[0.65773, 0.6578]$ in panel (a).
    (c) Spectrum as a function of $\Delta\beta$ at
    $C_i=0.248$.
    (d) Enlarged view of a narrow interval $[3.4277, 3.428]$ in panel (c).
    The successive emergence of subharmonic branches in panels
    (b) and (d) indicates period-doubling-like fine structures.
    For visualization, panels (a) and (c) were constructed by grouping every 20 adjacent Fourier bins and retaining the maximum power within each group, whereas panels (b) and (d) used the same procedure with groups of five bins. 
    Each grouped value was plotted at the mean frequency of its constituent bins.
    }
    
    \label{fig:frequency_spectra}
\end{figure}

This symmetry motivates the use of the modulus of the order parameter,
\begin{equation}
R(t)=|Z(t)|,
\label{eq:observable_R}
\end{equation}
as a bifurcation observable. 
Under a global rotation, $Z\rightarrow e^{\mathrm{i}\psi}Z$, whereas $|Z|\rightarrow|Z|$. 
Therefore, $R(t)$ is invariant under the trivial collective rotation and serves as a scalar observable of the internal amplitude and relative-phase dynamics.
For a phase-locked limit cycle, $R(t)$ is constant, whereas quasiperiodic and chaotic attractors produce periodic, quasiperiodic, or irregular variations of $R(t)$.

We characterize these variations using the Hann-windowed discrete Fourier transform. 
For a post-transient segment $x_n=R(t_n)$, $n=0,\ldots,N-1$, we define
\begin{equation}
Y_k
=
\sum_{n=0}^{N-1}
\left(x_n-\langle x \rangle\right)h_n
\exp\left(-2\pi\mathrm{i}\frac{nk}{N}\right),
\end{equation}
where
\begin{equation}
\langle x \rangle
=
\frac{1}{N}\sum_{n=0}^{N-1}x_n
\end{equation}
is the segment mean, and
\begin{equation}
h_n=
\frac{1}{2}
\left[
1-\cos\left(\frac{2\pi n}{N-1}\right)
\right]
\end{equation}
is the Hann window. 
The $k$th Fourier bin corresponds to the frequency
\begin{equation}
f_k=\frac{k f_s}{N},
\end{equation}
where $f_s=100$ is the sampling frequency.
For the one-sided spectrum, a factor of $2$ is applied to $0<k<N/2$ :
\begin{equation}
P_k
=
\frac{2|Y_k|^2}{f_s\sum_{n=0}^{N-1}h_n^2}.
\label{eq:PSD_R}
\end{equation}
We divided the post-transient time series into 16 segments of length $N=2^{20}$, with 50\% overlap between consecutive segments. 
The resulting power spectra were averaged over the segments and are denoted by $\langle P_k\rangle$ in Figure~\ref{fig:frequency_spectra}.

The phase-locked limit cycle has no appreciable spectral component at nonzero frequency because $R(t)$ is constant. 
A two-dimensional torus contains the global rotation together with one additional frequency. 
Since the global rotation is removed by taking $|Z|$, the spectrum of $R(t)$ contains one fundamental frequency and its harmonics,
\begin{equation}
f_n=n f_0,
\qquad
n=1,2,\ldots .
\end{equation}
The 2-torus is therefore identified in the spectral maps by a family of discrete, approximately equally spaced frequency branches. 
In contrast, chaotic dynamics produces densely distributed spectral components and
a broadband background.

Figures~\ref{fig:frequency_spectra}(a) and \ref{fig:frequency_spectra}(c) show the spectral evolution along the two cuts in Fig.~\ref{fig:bifurcation_landscape}. 
Although $C_i$ and $\Delta\beta$ have opposite dynamical effects, the two maps exhibit similar spectral structures. 
Increasing $\Delta\beta$ destabilizes the phase-locked state and leads toward chaos, whereas increasing $C_i$ at fixed heterogeneity regularizes the dynamics and eventually restores phase locking.

The enlarged views in Figs.~\ref{fig:frequency_spectra}(b) and
\ref{fig:frequency_spectra}(d) reveal repeated generation of
subharmonic branches over narrow parameter intervals. 
The similar branch-splitting structures along both cuts suggest that the heterogeneity-driven transition to chaos and the coupling-driven reverse transition pass through closely related intermediate attractors. 
These fine structures are consistent with period-doubling-like bifurcations of the amplitude dynamics toward chaos.

\begin{figure}[t]
    \centering
    \includegraphics[width=0.92\linewidth]
    {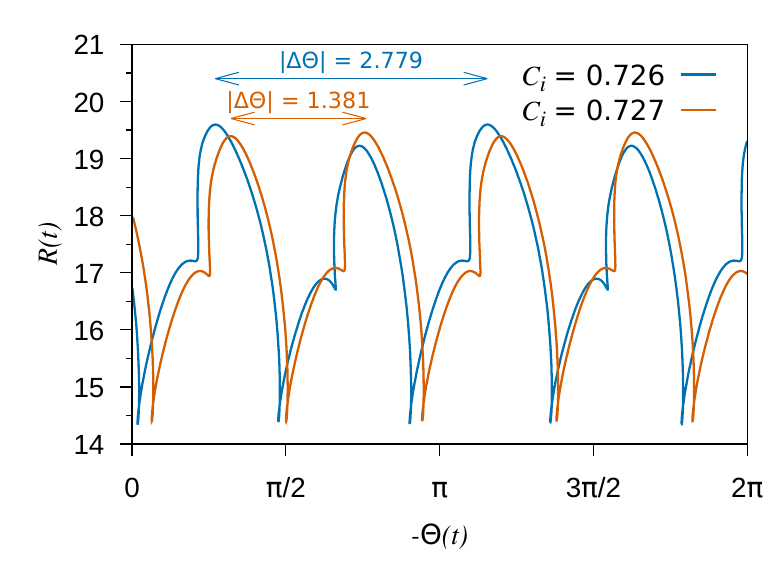}
    \caption{
    Geometric signature of a period-doubling-like transition.
    The order-parameter amplitude $R(t)$ is plotted against the reversed
    phase $-\Theta(t)$ for $C_i=0.726$ and $C_i=0.727$.
    At $C_i=0.727$, a single amplitude cycle repeats after a phase advance
    $|\Delta\Theta|=1.381$.
    At $C_i=0.726$, neighboring high and low peaks alternate, and two
    amplitude cycles are required to recover the same pattern, corresponding
    to a total phase advance $|\Delta\Theta|=2.779$, approximately twice that
    of the single cycle.
    }
    \label{fig:Z_phase_amplitude_PD}
\end{figure}

This interpretation is further supported by the amplitude--phase
portrait of the order parameter $Z$ in Fig.~\ref{fig:Z_phase_amplitude_PD}. 
Due to the global $U(1)$ equivariance, a dynamically rotating state may return to the same amplitude profile after one rotational period while the collective phase advances by a fixed amount. 
Thus, one amplitude cycle is associated with a well-defined collective phase increment. 
Comparing the two nearby parameter values, the $C_i=0.727$ trajectory displays a single-cycle pattern, whereas the $C_i=0.726$ trajectory shows two successive amplitude peaks of different heights. 
This means that the former amplitude cycle is split into a doubled cycle: the trajectory must pass through two alternating amplitude excursions before recovering the same dynamical pattern. 
Accordingly, the phase advance associated with each excursion, $|\Delta\Theta|=1.381$, is approximately halved relative to the full doubled cycle, $|\Delta\Theta|=2.779$.
The alternating amplitude peaks and the approximately doubled phase advance required to recover the same dynamical pattern provide a geometric signature consistent with a period-doubling scenario.

\section{Polygonal manifold featuring $Z=0$}
\label{sec:polygonal_manifold}

Beyond the dynamical regimes displayed in Fig.~\ref{fig:bifurcation_landscape}, we next uncover a special family of collective states characterized by a vanishing order parameter, $Z=0$, which is not represented in the single-initial-condition regime map.

Substituting $Z=0$ into Eq.~\eqref{eq:model_mean_field} gives
\begin{equation}
\dot{z}_j
=
\left[
\beta_j-MC_r-k_r|z_j|^2
+
\mathrm{i}
\left(
\omega-MC_i
\right)
\right]z_j.
\label{eq:zero_Z_dynamics}
\end{equation}
Each oscillator with
$\beta_j>MC_r$ therefore admits a constant-amplitude solution
\begin{equation}
r_j^{*}
=
\sqrt{\frac{\beta_j-MC_r}{k_r}},
\label{eq:center_amplitude}
\end{equation}
whereas oscillators with $\beta_j\leq MC_r$ remain at $z_j=0$.
All active oscillators rotate with the same angular frequency
\begin{equation}
\Omega^{*}=\omega-MC_i.
\label{eq:center_frequency}
\end{equation}
The corresponding solutions can be written as
\begin{equation}
z_j(t)
=
r_j^{*}
e^{\mathrm{i}(\Omega^{*} t+\theta_j^{*})},
\label{eq:center_solution}
\end{equation}
where the constant phases $\theta_j^{*}$ satisfy
\begin{equation}
\sum_{j\in\mathcal{A}}
r_j^{*}e^{\mathrm{i}\theta_j^{*}}=0.
\label{eq:polygon_closure}
\end{equation}
Here,
\begin{equation}
\mathcal{A}
=
\left\{
j\mid\beta_j>MC_r
\right\}
\end{equation}
denotes the set of active oscillators, and $N_a=|\mathcal{A}|$ is their number.

Equation~\eqref{eq:polygon_closure} also proves that the solutions constructed from Eq.~\eqref{eq:center_solution} remain on the $Z=0$ for all time.
Indeed, using Eq.~\eqref{eq:center_solution}, one obtains
\begin{equation}
Z(t)
=
\frac{e^{\mathrm{i}\Omega^{*} t}}{M}
\sum_{j\in\mathcal{A}}
r_j^{*}e^{\mathrm{i}\theta_j^{*}}
=
0,
\end{equation}
and hence
\begin{equation}
\dot{Z}
=
\mathrm{i}\Omega^{*} Z
=
0.
\label{eq:Z_invariance}
\end{equation}
Thus, every set of phases satisfying Eq.~\eqref{eq:polygon_closure} defines a phase-locked cycle with fixed amplitudes and fixed relative phases.

Geometrically, Eq.~\eqref{eq:polygon_closure} requires the complex vectors $r_j^{*}e^{\mathrm{i}\theta_j^{*}}$ to form a closed polygon whose side lengths are fixed by Eq.~\eqref{eq:center_amplitude}. 
The existence of such a configuration requires the longest side not to exceed the sum of the remaining side lengths. 
For a generic polygon with $N_a$ fixed side lengths, the configuration is specified by $N_a$ angular variables. 
The closure condition for a closed polygon imposes two independent real constraints. 
Therefore, the full solution manifold has dimension
\begin{equation}
d_{\mathrm{manifold}}
=
N_a-2 .
\label{eq:center_dimension}
\end{equation}
This manifold still includes the common rotational degree of freedom.
After quotienting out the global rotation, the remaining shape space has dimension
\begin{equation}
d_{\mathrm{shape}}
=
d_{\mathrm{manifold}}-1
=
N_a-3 .
\label{eq:shape_dimension}
\end{equation}
For $N_a\geq4$, this manifold is therefore composed of infinitely many one-dimensional phase-locked cycles parameterized by a $d_{\mathrm{shape}}$-dimensional space of polygonal shapes.

We next examine the transverse stability of the periodic solutions. Introducing the co-rotating variables
\begin{equation}
w_j(t)=z_j(t)e^{-\mathrm{i}\Omega^{*}t},
\end{equation}
transforms each periodic solution in Eq.~\eqref{eq:center_solution} into a fixed point,
\begin{equation}
w_j^{*}=r_j^{*}e^{\mathrm{i}\theta_j^{*}}.
\label{fixed_points}
\end{equation}
Let $\boldsymbol{q}^{*}$ denote the corresponding $2M$-dimensional real state. 
In the co-rotating frame, the vector field is
\begin{equation}
\boldsymbol{F}_{\mathrm{rot}}(\boldsymbol{q})
=
\boldsymbol{F}(\boldsymbol{q})
-
\Omega^{*}\mathcal{G}\boldsymbol{q},
\end{equation}
where $\mathcal{G}$ is the generator of the global phase rotation. 
Linear perturbations around $\boldsymbol{q}^{*}$ obey
\begin{equation}
\delta\dot{\boldsymbol{q}}
=
J_{\mathrm{rot}}(\boldsymbol{q}^{*})
\delta\boldsymbol{q},
\end{equation}
where
\begin{equation}
J_{\mathrm{rot}}(\boldsymbol{q}^{*})
=
\left.
\frac{\partial\boldsymbol{F}_{\mathrm{rot}}}
{\partial\boldsymbol{q}}
\right|_{\boldsymbol{q}=\boldsymbol{q}^{*}}.
\label{eq:center_jacobian}
\end{equation}
Thus, the Floquet stability of the original periodic orbit is determined by the eigenvalues of the co-rotating Jacobian around the fixed point.

Rotational equivariance implies that two globally rotated fixed points are related by
\begin{equation}
\boldsymbol{q}_{\psi}^{*}
=
G_{\psi}\boldsymbol{q}^{*},
\end{equation}
and their Jacobians satisfy
\begin{equation}
J_{\mathrm{rot}}(\boldsymbol{q}_{\psi}^{*})
=
G_{\psi}
J_{\mathrm{rot}}(\boldsymbol{q}^{*})
G_{\psi}^{-1}.
\label{eq:jacobian_similarity}
\end{equation}
Therefore, the Jacobians are similar matrices and all fixed points corresponding to a periodic solution Eq.~\eqref{eq:center_solution} have identical eigenspectra, making it sufficient to evaluate the Jacobian at a single representative point in the $d_{\mathrm{shape}}$-dimensional polygonal shape space. 

\begin{figure}[t]
    \centering
    \includegraphics[width=0.9\linewidth]
    {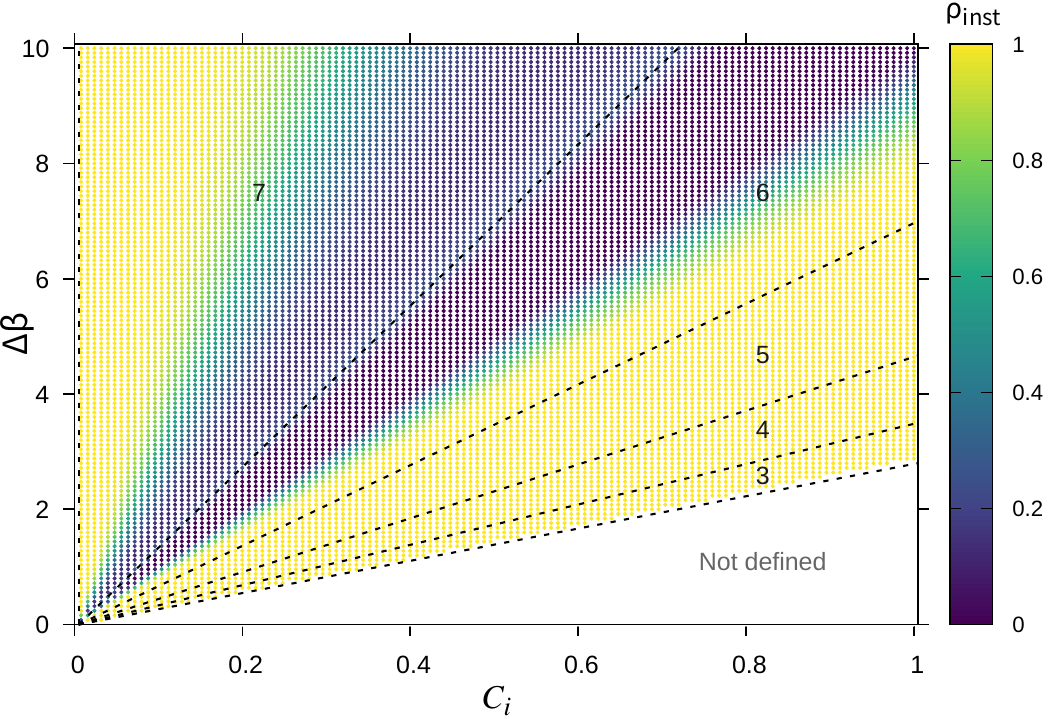}
    \caption{
    Sampled instability ratio of the polygonal manifold in
    the $(C_i,\Delta\beta)$ plane. The color represents the fraction
    $\rho_{\mathrm{inst}}$ of sampled polygonal configurations that are
    transversely unstable. Blue corresponds to a predominantly stable
    manifold and yellow to a predominantly unstable manifold.
    The dashed lines mark the parameter values at which an additional
    oscillator becomes active, $\beta_j=MC_r$, and therefore separate
    polygonal solutions with different numbers $N_a$ of nonzero sides.
    The integers assigned in different parameter regions correspond to $N_a$.
    The blank region corresponds either to $N_a<3$ or to active amplitudes that do not satisfy the polygon closure condition; in either case, the polygonal solution is not defined.
    }
    \label{fig:center_manifold_stability}
\end{figure}

To analyze the stability of these representative points, we sample the polygonal shape space spanned by the $N_a-3$ independent variables $\theta_j^{*}$ with $M-N_a+2\leq j\leq M-2$ chosen from uniformly distributed random numbers in $[0,2\pi]$.
The global rotational degree of freedom is removed by fixing $\theta_{M-N_a+1}^{*}=0$. 
For each sampled point, the remaining two phases, $\theta_j^{*}$ with $M-2<j$, are determined by numerically solving Eq.~\eqref{eq:polygon_closure} using Newton's method. 
A proposal is accepted when the norm of the polygon-closure residual is below $10^{-12}$; 
otherwise, it is discarded and resampled. 
Sampling continues until the prescribed number of polygonal configurations, $N_{\mathrm{sample}}$, has been accepted, subject to a maximum of $100N_{\mathrm{sample}}$ attempts.
In the numerical data used for Fig.~\ref{fig:center_manifold_stability}, every geometrically admissible parameter point reached the prescribed sample size, and no sampling-incomplete point was found.
A sampled cycle is classified as transversely unstable when at least one eigenvalue outside these neutral directions has a positive real part. 
In the numerical implementation, ``positive'' means
\begin{equation}
\max_{\lambda\in\sigma(J_{\mathrm{rot}})}
\operatorname{Re}\lambda
>
\varepsilon_{\mathrm{eig}},
\qquad
\varepsilon_{\mathrm{eig}}
=
10^{-6}\lVert J_{\mathrm{rot}}\rVert_{\mathrm{F}},
\end{equation}
where $\lVert\cdot\rVert_{\mathrm{F}}$ denotes the Frobenius norm. 
This scale-dependent threshold prevents small numerical deviations of the theoretically neutral eigenvalues from being misidentified as transverse instability.
Therefore, the sampled instability ratio, defined as the fraction of transversely unstable configurations obtained under this sampling protocol,
is given by
\begin{equation}
\rho_{\mathrm{inst}}
=
\frac{N_{\mathrm{unstable}}}{N_{\mathrm{sample}}},
\label{eq:instability_ratio}
\end{equation}
where
\begin{equation}
N_{\mathrm{sample}}
=
\begin{cases}
1, & d_{\mathrm{shape}}=0,\\[2mm]
d_{\mathrm{shape}}\times 10^3, & d_{\mathrm{shape}}>0,
\end{cases}
\end{equation}
and $N_{\mathrm{unstable}}$ denotes the number of transversely unstable ones.
Thus, $\rho_{\mathrm{inst}}=0$ indicates that all sampled cycles are stable, whereas $\rho_{\mathrm{inst}}=1$ indicates that all of them are unstable.

In the present work, the Jacobians of a polygonal configuration and its global complex-conjugate mirror in terms of the augmented perturbation vector $(\delta\boldsymbol{w},\delta\boldsymbol{w}^{*})^{\mathrm{T}}$ satisfy
\begin{equation}
J_{\mathrm{mirror}}=J^{\mathrm{T}}.
\end{equation}
Since matrix transposition leaves the characteristic polynomial, and hence the eigenvalue spectrum, unchanged, the two configurations are isospectral. 
Consequently, the global mirror of an accepted configuration has identical transverse stability and is not added as a separate independent sample.

Figure~\ref{fig:center_manifold_stability} presents the instability ratio $\rho_{\mathrm{inst}}$ of the polygonal manifold over the same $(C_i,\Delta\beta)$ parameter plane as Fig.~\ref{fig:bifurcation_landscape}.
The dashed lines divide the parameter plane into triangular, quadrilateral, pentagonal, hexagonal, and heptagonal solution sectors, each assigned an integer equal to the $N_a$. 
Because the amplitudes are given by Eq.~\eqref{eq:center_amplitude}, these boundaries occur whenever
\begin{equation}
\beta_j=MC_r,
\end{equation}
at which one polygon side continuously appears or disappears.

The triangular, quadrilateral, and pentagonal sectors exhibit large instability ratios, indicating that most sampled cycles are transversely unstable and helping explain why the prescribed initial condition does not select polygonal cycles in these sectors.
In the triangular sector, the unique polygonal cycle is transversely unstable and cannot be selected as an attractor. 
In the quadrilateral and pentagonal sectors, any stable polygonal states selected by this initial condition would have two and three neutral directions, respectively, and would therefore be misclassified as two- and three-tori based solely on the Lyapunov spectrum.
However, the corresponding torus regions in Fig.~\ref{fig:bifurcation_landscape} lie outside these sectors, confirming that zero Lyapunov exponents of the observed torus attractors do not originate from neutral directions tangent to the polygonal manifold.
In contrast, a broad region spanning the hexagonal and heptagonal sectors has a small instability ratio, implying that a substantial fraction of the sampled configurations are stable. 
Unfortunately, none of these stable polygonal states is selected by the prescribed initial condition, as evidenced by the absence of parameter points with four or more zero Lyapunov exponents in Fig.~\ref{fig:bifurcation_landscape}.
The stability of the polygonal manifold is therefore governed not only by the existence or dimensionality of the solution set, but also by the specific polygonal geometry determined by the side lengths and relative angles.

This configuration dependence has direct consequences for the global dynamics. 
At the same parameter values, stable and unstable polygonal cycles may coexist even though they share the same oscillator amplitudes and collective frequency. 
Stable portions of the polygonal manifold can therefore coexist with other attractors and provide an additional source of multistability. 
Compared with the single-initial-condition regime map in Fig.~\ref{fig:bifurcation_landscape}, the stable regions in Fig.~\ref{fig:center_manifold_stability} suggest that sampling additional
initial conditions may reveal finite basins of attraction associated with the polygonal manifold. 
This possibility is examined systematically through the Lyapunov-exponent distributions in the following section.

\section{Multistability revealed by Lyapunov-exponent distributions}
\label{sec:multistability}

The dynamical-regime diagram in Fig.~\ref{fig:bifurcation_landscape} shows only the attractor reached from one prescribed initial condition, based on which the spectra in Fig.~\ref{fig:frequency_spectra} are obtained. 
Therefore, the patchy interspersion of dynamical regimes near the boundaries of the chaotic region in Fig.~\ref{fig:bifurcation_landscape}, together with the abrupt jumps between spectral branches in Fig.~\ref{fig:frequency_spectra}, may partly reflect basin-dependent attractor selection in multistable parameter regions, rather than a continuous bifurcation of a single attractor.
Moreover, the stable portions of the polygonal manifold identified above provide an additional family of coexisting attractors, further enriching the multistable dynamics.

To quantify coexistence, we compute the largest Lyapunov exponent for
\begin{equation}
N_{\mathrm{s}}=1000
\end{equation}
random initial conditions at each parameter value $p$, representing either $C_i$ or $\Delta\beta$. 
For each oscillator, the real and imaginary parts of the initial state are sampled independently and uniformly from $[-5,5]$, such that $z_j(0)\in[-5,5]\times[-5,5]$ in the complex plane.
Let
\begin{equation}
\lambda_{\max}^{(s)}(p),
\qquad
s=1,\ldots,N_{\mathrm{s}},
\end{equation}
be the largest Lyapunov exponent obtained from the $s$th initial
condition. 
We then construct a normalized histogram of these exponents. 
Let $\mathcal{B}_k$ be the $k$th bin of largest Lyapunov exponents. 
The empirical probability that $\lambda_{\max}$ falls into this bin is
\begin{equation}
\rho_k(p)
=
\frac{1}{N_{\mathrm{s}}}
\sum_{s=1}^{N_{\mathrm{s}}}
\mathbb{I}
\left[
\lambda_{\max}^{(s)}(p)\in\mathcal{B}_k
\right],
\end{equation}
where
\begin{equation}
\mathbb{I}
\left[
\lambda_{\max}^{(s)}(p)\in\mathcal{B}_k
\right]
=
\begin{cases}
1, & \lambda_{\max}^{(s)}(p)\in\mathcal{B}_k,\\
0, & \lambda_{\max}^{(s)}(p)\notin\mathcal{B}_k.
\end{cases}
\end{equation}
Thus, $\rho_k(p)$ measures the basin probability of trajectories whose
largest Lyapunov exponent lies in the bin $\mathcal{B}_k$.

We classify a trajectory as chaotic when
\begin{equation}
\lambda_{\max}^{(s)}(p)\geq
\varepsilon_{\mathrm{LE}}.
\end{equation}
States satisfying $\lambda_{\max}^{(s)}(p)<\varepsilon_{\mathrm{LE}}$ are further classified using finite-time frequency-locking, amplitude-stationarity, and order-parameter criteria. 
Each trajectory is monitored over an additional time window $t_m=2000$. 
At every sampled time $t_n$, we define the instantaneous mean amplitude
\begin{equation}
\overline{r}(t_n)
=
\frac{1}{M}\sum_{j=1}^{M}r_j(t_n)
\end{equation}
and regard oscillator $j$ as phase active when
\begin{equation}
r_j(t_n)
>
\varepsilon\left[1+\overline{r}(t_n)\right],
\qquad
\varepsilon=10^{-6}.
\label{eq:active_oscillator}
\end{equation}
For each active oscillator, the instantaneous angular frequency is evaluated directly from the vector field as
\begin{equation}
\dot{\theta}_j(t_n)
=
\frac{
\operatorname{Im}\!\left[
z_j^{*}(t_n)\dot{z}_j(t_n)
\right]
}{
r_j^2(t_n)
}.
\end{equation}
At each sampled time with at least two active oscillators, let $\dot{\theta}_{\min}(t_n)$ and $\dot{\theta}_{\max}(t_n)$ denote the minimum and maximum instantaneous frequencies among them. 
Frequency locking requires
\begin{equation}
\dot{\theta}_{\max}(t_n)
-
\dot{\theta}_{\min}(t_n)
\leq
\varepsilon
\left[
1+
\max\left(
|\dot{\theta}_{\min}(t_n)|,
|\dot{\theta}_{\max}(t_n)|
\right)
\right]
\label{eq:freq_locking}
\end{equation}
at every sampled time $t_n$ in the monitoring window.

Amplitude stationarity is tested using the full-window extrema
\begin{equation}
r_j^{\min}
=
\min_{t_n\in t_m}r_j(t_n),
\qquad
r_j^{\max}
=
\max_{t_n\in t_m}r_j(t_n).
\end{equation}
The amplitude of every oscillator is regarded as stationary when
\begin{equation}
r_j^{\max}-r_j^{\min}
\leq
\varepsilon\left(1+r_j^{\max}\right),
\qquad
j=1,\ldots,M .
\label{eq:amp_stationarity}
\end{equation}
A trajectory satisfying both Eqs.~\eqref{eq:freq_locking} and
\eqref{eq:amp_stationarity} is classified as locked. 
Otherwise, it is classified as an unlocked state, predominantly associated with quasiperiodic tori.

For locked states, we further distinguish ordinary phase-locked cycles from the vanishing-order-parameter cycles discussed in Sec.~\ref{sec:polygonal_manifold}. 
A vanishing-order-parameter state is identified when
\begin{equation}
|Z(t_n)|
\leq
\varepsilon\left[1+\overline{r}(t_n)\right]
\label{eq:zero_order_parameter_classification}
\end{equation}
at every sampled time. 

No fixed-point attractors were numerically detected among the sampled trajectories.
Accordingly, trajectories with $\lambda_{\max}<\varepsilon_{\mathrm{LE}}$ are treated as zero-Lyapunov states and divided into three classes:
\begin{equation}
\begin{aligned}
\alpha=1 &: \quad |Z|>0 \ \text{phase-locked cycle},\\
\alpha=2 &: \quad |Z|=0 \ \text{phase-locked cycle on the polygonal manifold},\\
\alpha=3 &: \quad \text{unlocked states including tori}.
\end{aligned}
\end{equation}
The basin probability of each class is
\begin{equation}
P_{\alpha}(p)
=
\frac{N_{\alpha}(p)}{N_{\mathrm{s}}}.
\end{equation}

\begin{figure}[t]
    \centering
    \includegraphics[width=0.85\linewidth]
    {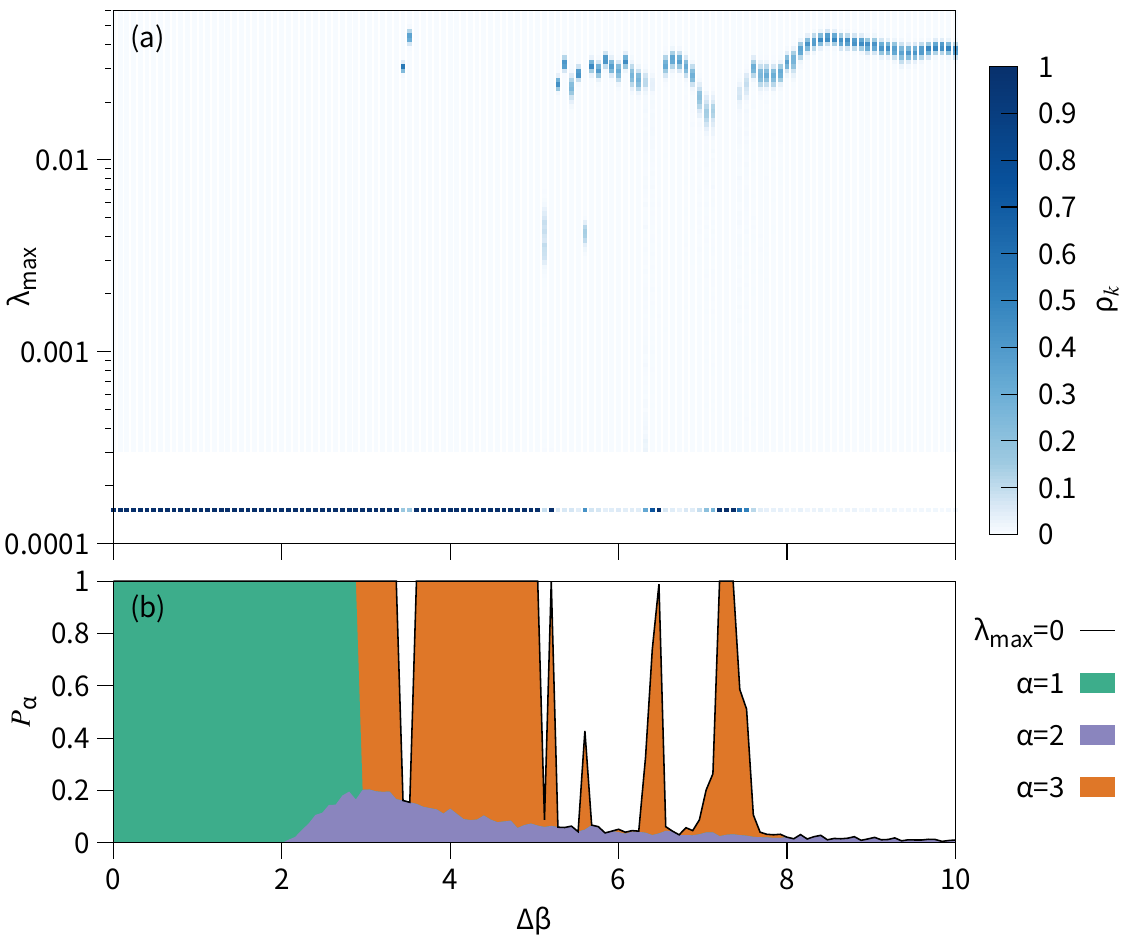}
    \caption{
    Multistability as a function of heterogeneity at $C_i=0.248$.
    (a) Distribution of the largest Lyapunov exponent over $N_{\mathrm{s}}=1000$ random initial conditions, shown on a logarithmic vertical axis.
    All estimates satisfying $\lambda_{\max}<\varepsilon_{\mathrm{LE}}$ are collected into a single subthreshold bin plotted at $1.5\times10^{-4}$.
    (b) Basin probabilities $P_{\alpha}$ of zero-Lyapunov states, decomposed into $\alpha=1$ phase-locked cycles with nonzero order parameter, $\alpha=2$ phase-locked cycles with vanishing order parameter, and $\alpha=3$ unlocked states including tori. The total probability of zero-Lyapunov states $\lambda_{\max}=0$ is depicted with a black contour line. 
    }
    \label{fig:multistability_beta}
\end{figure}

Figure~\ref{fig:multistability_beta} shows that increasing
$\Delta\beta$ produces broad coexistence. 
At small heterogeneity, most trajectories converge to ordinary phase-locked cycles.
As $\Delta\beta$ increases, the polygonal manifold acquires finite basins, consistent with the instability ratio found in Fig.~\ref{fig:center_manifold_stability}. 
At larger heterogeneity, positive Lyapunov exponents appear while zero-Lyapunov states persist, showing coexistence among chaotic attractors, the polygonal manifold and tori.

\begin{figure}[t]
    \centering
    \includegraphics[width=0.85\linewidth]
    {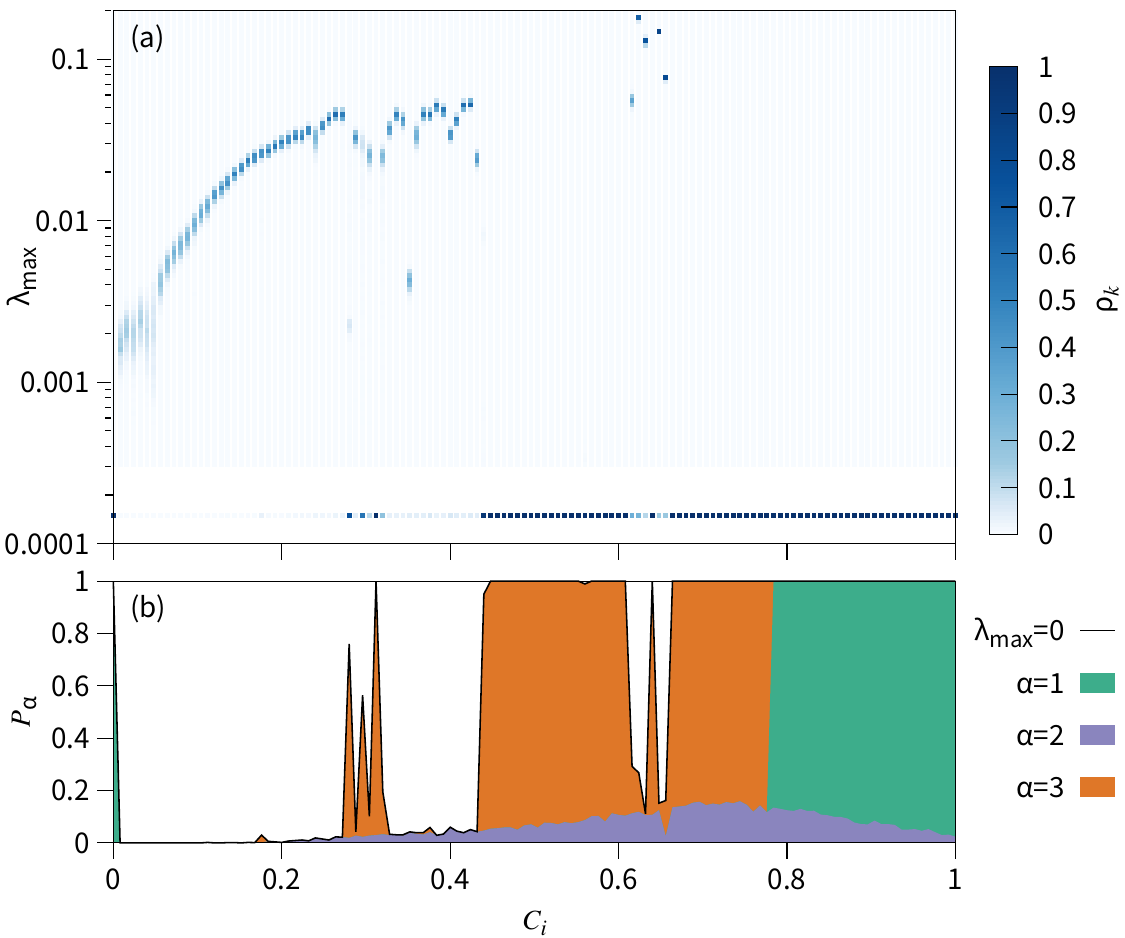}
    \caption{
    Multistability as a function of coupling at $\Delta\beta=9.12$.
    (a) Distribution of the largest Lyapunov exponent over $N_{\mathrm{s}}=1000$ random initial conditions, shown on a logarithmic vertical axis.
    All estimates satisfying $\lambda_{\max}<\varepsilon_{\mathrm{LE}}$ are collected into a single subthreshold bin plotted at $1.5\times10^{-4}$.
    (b) Basin probabilities $P_{\alpha}$ of zero-Lyapunov states, decomposed into $\alpha=1$ phase-locked cycles with nonzero order parameter, $\alpha=2$ phase-locked cycles with vanishing order parameter, and $\alpha=3$ unlocked states including tori. The total probability of zero-Lyapunov states $\lambda_{\max}=0$ is depicted with a black contour line.
    }
    \label{fig:multistability_Ci}
\end{figure}

Figure~\ref{fig:multistability_Ci} shows the reverse effect of the complex coupling. 
For weak coupling strength, chaotic states occupy a large fraction of the sampled basins. 
Stronger coupling $C_i$ eventually suppresses the positive-LLE basins and restores ordinary phase locking.
Together with the quasiperiodic branches already identified in Fig.~\ref{fig:frequency_spectra}, this supports the coexistence of chaos and tori in the same parameter range around $C_i=0.3$.

The similarity of the bifurcation structures and multistability patterns observed along the two parameter cuts in Fig.~\ref{fig:frequency_spectra} and Figs.~\ref{fig:multistability_beta} and \ref{fig:multistability_Ci}, together with the nearly linear regime boundaries in Figs.~\ref{fig:bifurcation_landscape} and nearly linear stability boundaries in Fig.~\ref{fig:center_manifold_stability}, can be understood from an approximate scaling of Eq.~\eqref{eq:model}. 
In the frame rotating with the common natural frequency, $w_j=z_j e^{-\mathrm{i}\omega t}$, and with $k_i=0$, we introduce the rescaled time
\begin{equation}
\tau=|C|t,
\qquad
|C|=\sqrt{C_r^2+C_i^2}.
\end{equation}
For $|C|>0$, Eq.~\eqref{eq:model} then becomes
\begin{equation}
\frac{\mathrm{d}w_j}{\mathrm{d}\tau}
=
\left[
\frac{\beta_j}{|C|}
-
\frac{k_r}{|C|}|w_j|^2
\right]w_j
+
\frac{C_r+\mathrm{i}C_i}{|C|}
\sum_{\ell\ne j}(w_\ell-w_j).
\label{eq:scaled_model}
\end{equation}
We further introduce the rescaled amplitude
\begin{equation}
u_j
=
\sqrt{\frac{k_r}{|C|}}\,w_j.
\label{eq:scaled_amplitude}
\end{equation}
Substitution into Eq.~\eqref{eq:scaled_model} gives
\begin{equation}
\frac{\mathrm{d}u_j}{\mathrm{d}\tau}
=
\left[
\frac{\beta_j}{|C|}
-
|u_j|^2
\right]u_j
+
\frac{C_r+\mathrm{i}C_i}{|C|}
\sum_{\ell\ne j}(u_\ell-u_j).
\label{eq:amplitude_scaled_model}
\end{equation}
This invertible amplitude-rescaling transformation, applied uniformly to all oscillators, preserves the qualitative attractor types and their stability while absorbing the coefficient $k_r/|C|$ exactly.
Thus, $k_r$ determines only the conversion between the physical and rescaled amplitudes and does not constitute an additional dimensionless control parameter.
Under the fixed coupling ratio in Eq.~\eqref{eq:coupling_ratio},
\begin{equation}
|C|
=
\frac{\sqrt{17}}{4}C_i,
\qquad
\frac{C_r+\mathrm{i}C_i}{|C|}
=
\frac{1+4\mathrm{i}}{\sqrt{17}},
\end{equation}
so that the normalized coupling term is independent of $C_i$. 
Using Eq.~\eqref{eq:beta_distribution}, Eq.~\eqref{eq:amplitude_scaled_model} becomes
\begin{equation}
\frac{\mathrm{d}u_j}{\mathrm{d}\tau}
=
\left[
\frac{\beta_{\min}}{|C|}
+
\frac{j-1}{M-1}\frac{\Delta\beta}{|C|}
-
|u_j|^2
\right]u_j
+
\frac{1+4\mathrm{i}}{\sqrt{17}}
\sum_{\ell\ne j}(u_\ell-u_j).
\label{eq:fully_scaled_model}
\end{equation}
Therefore, for fixed $M$ and coupling ratio, the remaining parameter dependence enters through the two dimensionless quantities $\beta_{\min}/|C|$ and $\Delta\beta/|C|$. 
Their relative magnitude is
\begin{equation}
\frac{\beta_{\min}}{\Delta\beta}.
\end{equation}
In the finite-heterogeneity regime relevant to the bifurcation structures considered here, $\Delta\beta\gg\beta_{\min}=0.01$, so that the contribution $\beta_{\min}/|C|$ can be treated as a subleading correction to the heterogeneity-dependent variation. 
To leading order, the parameter dependence is therefore approximately captured by the single ratio
\begin{equation}
\frac{\Delta\beta}{|C|}
=
\frac{4\Delta\beta}{\sqrt{17}C_i}.
\label{eq:scaled_control_parameter}
\end{equation}

This scaling provides a unified interpretation of the two-dimensional parameter maps. 
Parameter points with the same value of $\Delta\beta/|C|$ are expected to exhibit similar leading-order reduced dynamics, although they are not exactly dynamically equivalent.
Consequently, transitions between different dynamical regimes occur at approximately fixed values of $\Delta\beta/|C|$, implying
\begin{equation}
\Delta\beta \propto C_i,
\end{equation}
and thereby accounting for the nearly linear organization of the regime boundaries in Fig.~\ref{fig:bifurcation_landscape} and stability boundaries in Fig.~\ref{fig:center_manifold_stability}.
Moreover, since $C_i=4/\sqrt{17}$ gives $|C|=1$, the scaling maps each parameter point $(C_i,\Delta\beta)$ approximately to the point
\[
\left(\frac{4}{\sqrt{17}},\,\frac{\Delta\beta}{|C|}\right)
\]
on this vertical cut, preserving the effective control ratio $\Delta\beta/|C|$.
The $C_i$ and $\Delta\beta$ cuts therefore sample approximately the same underlying bifurcation structure through different parameterizations, explaining the similarity of their spectral sequences and multistability patterns.

The same scaling also accounts for the relative shift of the characteristic frequencies of $R(t)$. 
Two parameter points with the same value of $\Delta\beta/|C|$ are expected, within the above approximation, to follow similar reduced dynamics while evolving on different physical time scales. 
Since $\tau=|C|t$, their characteristic periods and frequencies satisfy
\begin{equation}
\frac{T_1}{T_2}
\simeq
\frac{|C_2|}{|C_1|},
\qquad
\frac{f_1}{f_2}
\simeq
\frac{|C_1|}{|C_2|}.
\end{equation}
This prediction applies directly to the spectra in Fig.~\ref{fig:frequency_spectra} and is consistent with the relative displacement of their spectral branches.

Because the initial conditions are specified within fixed domains in the original variables, the corresponding initial conditions in the rescaled phase space depend on $|C|$. 
In the dynamical-regime map Fig.~\ref{fig:bifurcation_landscape}, this dependence can alter basin selection along contours of constant $\Delta\beta/|C|$ in multistable regions, leading to deviations of the observed regime boundaries from the ideal linear relation $\Delta\beta\propto C_i$. 
It also prevents parameter points on the two cuts with the same value of $\Delta\beta/C_i$ from being initialized at the same rescaled states, so that their spectral sequences and multistability patterns need not agree pointwise. 
The scaling therefore predicts only a qualitative similarity among these numerical results.

In summary, the ensemble analysis reveals the coexistence of basins associated with chaotic attractors, quasiperiodic tori, phase-locked limit cycles, and the polygonal solution manifold. 
Although $\Delta\beta$ and $C_i$ represent distinct physical controls, their dominant competition is approximately organized by the effective ratio $\Delta\beta/|C|$, which compares the heterogeneity scale with the strength of the complex coupling. 
The two parameter cuts can therefore be interpreted as different projections of a common leading-order bifurcation structure governed primarily by this ratio, providing a unified explanation for the nearly linear regime boundaries, similar period-doubling-like spectral sequences, and corresponding multistability patterns.

\section{Discussion}
\label{sec:discussion}

We have systematically characterized the collective dynamics generated by the interplay between linear-growth-rate heterogeneity and complex global coupling in SL oscillators. 
Whereas quasiperiodicity and chaos in SL systems have commonly been associated with Kerr-type nonlinear frequency shifts or natural-frequency heterogeneity, in the present system they arise from heterogeneity in the preferred oscillation amplitudes.
The resulting mean-field mismatch is rotated by the complex coupling into a substantial radial component, whose competition with local nonlinear radial restoration destabilizes phase locking.
Using the rotationally invariant observable $R(t)=|Z(t)|$, we identify spectral and geometric signatures consistent with period-doubling-like transitions, including subharmonic branch splitting and an approximately doubled phase advance. 
Whereas Lee et al.~\cite{shing2013phase} analyzed the linear stability of an incoherent $Z=0$ state in the continuum limit $N\to\infty$, we construct a finite-size manifold of phase-locked $Z=0$ solutions with a closed-polygon geometry and quantify its configuration-dependent transverse stability through an instability ratio.
Ensemble sampling further reveals coexistence among polygonal phase-locked cycles, ordinary phase-locked cycles, quasiperiodic tori, and chaotic attractors.
The approximate scaling suggests that the obtained results are organized by the ratio $\Delta\beta/|C|$, thereby highlighting the competition between heterogeneity and complex coupling.

The present analysis is restricted to a finite system with $M=8$ and an evenly spaced distribution of linear growth rates. 
For larger $M$, the radial and angular balance conditions, Eqs.~\eqref{static balance radial} and~\eqref{static balance angular}, must be satisfied simultaneously by a greater number of oscillators, while the collective mean field and the dimensionality of the dynamics also change.
The relative prevalence of phase-locked, quasiperiodic, and chaotic states therefore cannot be inferred directly from the present finite-size results.
Hence, extensions to larger populations and other heterogeneity distributions~\cite{shing2013phase,mirollo1990amplitude} are required to determine whether the identified mechanism persists in the thermodynamic limit and can support extensive chaos~\cite{hohlein2019lyapunov,wolfrum2011spectral} or cluster-mediated dynamics~\cite{shing2013phase,nakagawa1994collective}. 
The fixed ratio $C_r/C_i=0.25$ was selected because it provides a broad parameter window containing phase-locked, quasiperiodic, and chaotic regimes, thereby allowing their transitions and coexistence to be examined within a common parameter plane.
A systematic investigation of how the dynamical regimes depend on the coupling ratio, or equivalently on the argument of the complex coupling coefficient, and how this dependence changes with system size is left for future work.
Moreover, nonlinear~\cite{kuete2026synchronization,lee2022nontrivial}, nonlocal~\cite{premalatha2016imperfectly,zhao2018enhancing}, or nonreciprocal~\cite{ryu2026nonreciprocity,zhao2025nonreciprocal} coupling may provide additional routes to complex collective dynamics.

Despite these limitations, the present work shows that the bifurcation structures, polygonal solution manifold, and multistability patterns are interconnected consequences of the competition between heterogeneity and complex coupling. 
By systematically linking these structures, the present analysis establishes a foundation for studying complex collective dynamics in heterogeneous amplitude-inclusive oscillator networks.

\section*{Data availability}
The numerical data underlying the figures, together with the associated source code, data-processing and plotting scripts, and reproduction instructions, are publicly available on Zenodo at DOI: \href{https://doi.org/10.5281/zenodo.23181675}{10.5281/zenodo.23181675}.
\appendix
\section{Convergence of finite-time Lyapunov exponents}
\label{app:lle_convergence}

\begin{figure}[p]
\centering
\begin{minipage}{0.498\linewidth}
\begin{overpic}[width=\linewidth]{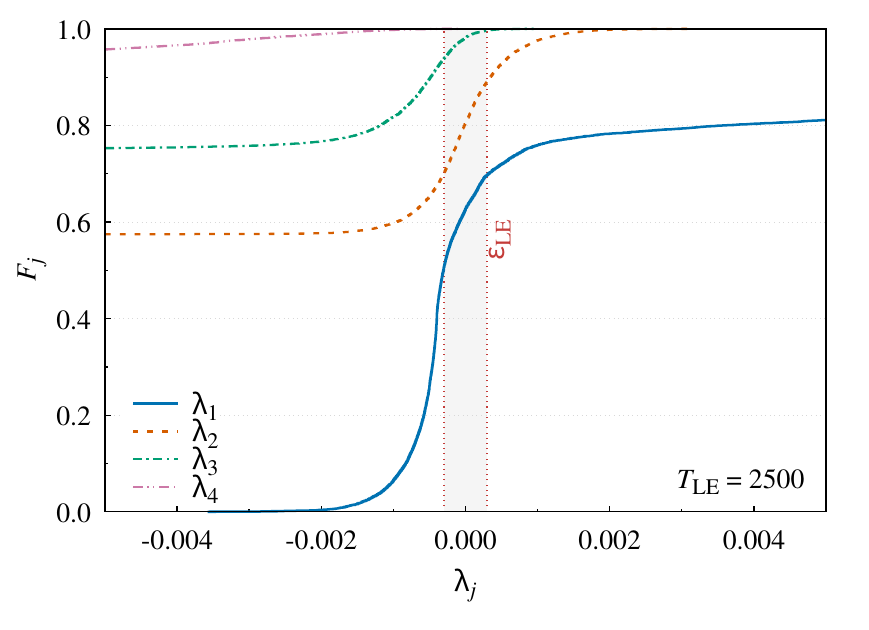}
    \put(13,62){\scriptsize\bfseries (a)}
\end{overpic}
\end{minipage}%
\begin{minipage}{0.498\linewidth}
\begin{overpic}[width=\linewidth]{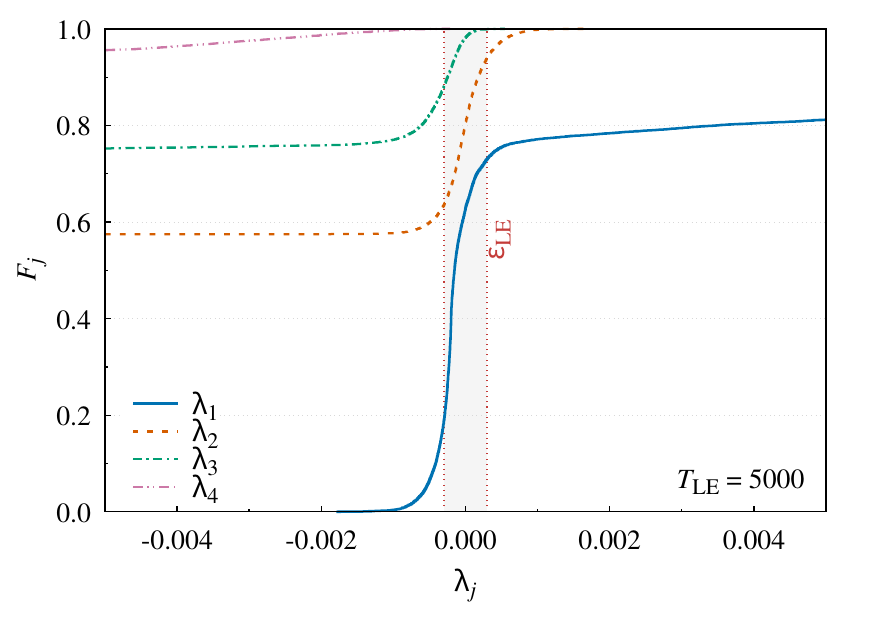}
    \put(13,62){\scriptsize\bfseries (b)}
\end{overpic}
\end{minipage}
\par\vspace{-1mm}

\begin{minipage}{0.498\linewidth}
\begin{overpic}[width=\linewidth]{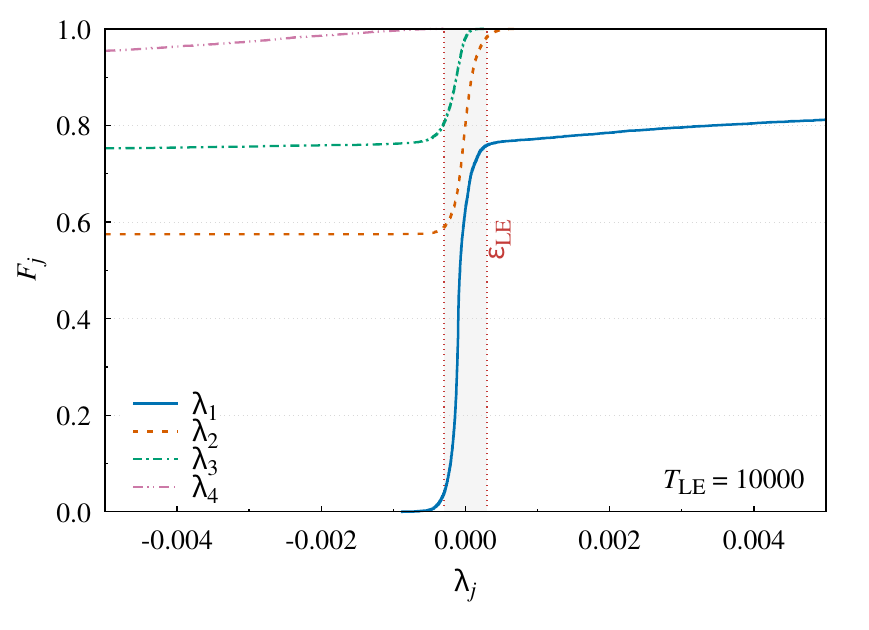}
    \put(13,62){\scriptsize\bfseries (c)}
\end{overpic}
\end{minipage}%
\begin{minipage}{0.498\linewidth}
\begin{overpic}[width=\linewidth]{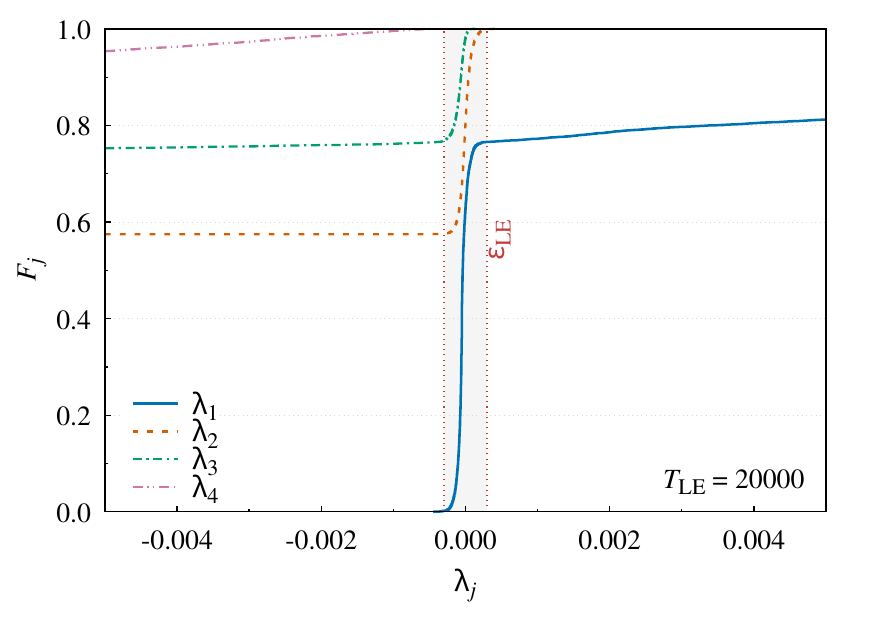}
    \put(13,62){\scriptsize\bfseries (d)}
\end{overpic}
\end{minipage}
\par\vspace{-1mm}

\begin{minipage}{0.498\linewidth}
\begin{overpic}[width=\linewidth]{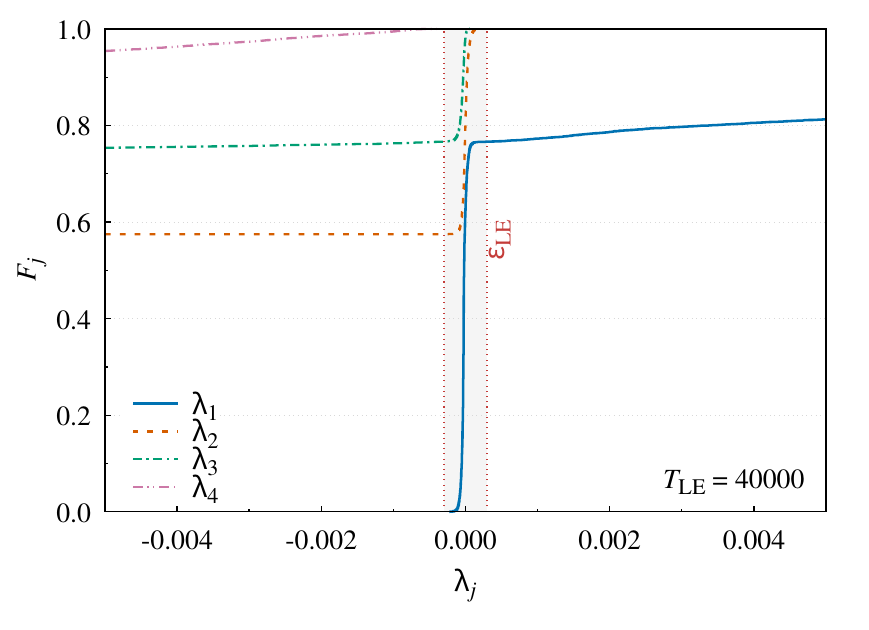}
    \put(13,62){\scriptsize\bfseries (e)}
\end{overpic}
\end{minipage}

\caption{
Empirical cumulative distribution functions of the four largest
finite-time Lyapunov exponents for
(a) $T_{\mathrm{LE}}=2500$,
(b) $T_{\mathrm{LE}}=5000$,
(c) $T_{\mathrm{LE}}=10000$,
(d) $T_{\mathrm{LE}}=20000$, and
(e) $T_{\mathrm{LE}}=40000$.
The shaded region denotes the numerical-zero interval
$|\lambda_j|\leq\varepsilon_{\mathrm{LE}}$.
The ECDFs sharpen with increasing $T_{\mathrm{LE}}$.
}
\label{fig:lle_convergence}
\end{figure}

The dynamical classification in Fig.~\ref{fig:bifurcation_landscape} uses the numerical-zero criterion $|\lambda_j|\leq\varepsilon_{\mathrm{LE}}$.
To assess its convergence, we compare the empirical cumulative distribution functions (ECDFs) of the four largest Lyapunov exponents for
$T_{\mathrm{LE}}=2500$, $5000$, $10000$, $20000$, and $40000$.
For the $j$th exponent, the ECDF is defined as
\begin{equation}
F_j(x;T_{\mathrm{LE}})
=
\frac{1}{N_p}
\sum_{q=1}^{N_p}
\mathbb{I}
\left[
\lambda_j^{(q)}(T_{\mathrm{LE}})\leq x
\right],
\end{equation}
where $N_p=15750$ is the number of sampled parameter points taken from the same $126\times126$ parameter grid in the $(C_i,\Delta\beta)$ plane in Fig.~\ref{fig:bifurcation_landscape}.
$\mathbb{I}[\cdot]$ denotes the indicator function, which equals 1 when the enclosed condition is satisfied and 0 otherwise.
The uncoupled boundary $C_i=0$ is excluded because its degenerate neutral directions are not representative of the coupled dynamical regimes.
Here, we use the prescribed initial condition.

The threshold $\varepsilon_{\mathrm{LE}}=3\times10^{-4}$ is chosen from the shoulders
of the central near-zero accumulation. 
The ECDFs rise rapidly where the finite-time estimates cluster around zero and become comparatively flat outside this interval, thereby separating the near-zero population from the positive and negative tails. 
The shoulder positions become stable for $T_{\mathrm{LE}}\geq20000$, supporting the adopted numerical-zero criterion $\varepsilon_{\mathrm{LE}}$ and the use of $T_{\mathrm{LE}}=40000$ in the dynamical classification.

\bibliographystyle{elsarticle-num} 
\bibliography{cas-refs}

@inproceedings{landau1944problem,
  title={On the problem of turbulence},
  author={Landau, Lev D},
  booktitle={Dokl. Akad. Nauk USSR},
  volume={44},
  pages={311},
  year={1944}
}

@article{stuart1960non,
  title={On the non-linear mechanics of wave disturbances in stable and unstable parallel flows Part 1. The basic behaviour in plane Poiseuille flow},
  author={Stuart, John Trevor},
  journal={Journal of Fluid Mechanics},
  volume={9},
  number={3},
  pages={353--370},
  year={1960},
  publisher={Cambridge University Press}
}

@incollection{kuramoto1984chemical,
  title={Chemical turbulence},
  author={Kuramoto, Yoshiki},
  booktitle={Chemical oscillations, waves, and turbulence},
  pages={111--140},
  year={1984},
  publisher={Springer}
}

@article{pikovsky2001synchronization,
  title={Synchronization},
  author={Pikovsky, Arkady and Rosenblum, Michael and Kurths, J{\"u}rgen},
  journal={Cambridge university press},
  volume={12},
  year={2001},
  publisher={Cambridge university press}
}

@article{kuete2026synchronization,
  title={Synchronization of nonlinearly coupled Stuart-Landau oscillators on networks},
  author={Kuete, Wilfried Segnou Tchio and Muolo, Riccardo and Dorchain, Marie and Nakao, Hiroya and Carletti, Teo},
  journal={Proceedings of the royal society A: mathematical, physical and engineering sciences},
  volume={482},
  year={2026},
  publisher={Royal Society Publishing}
}

@article{zhao2018enhancing,
  title={Enhancing coherence via tuning coupling range in nonlocally coupled Stuart--Landau oscillators},
  author={Zhao, Nannan and Sun, Zhongkui and Xu, Wei},
  journal={Scientific Reports},
  volume={8},
  number={1},
  pages={8721},
  year={2018},
  publisher={Nature Publishing Group UK London}
}

@article{zou2014emergence,
  title={Emergence of amplitude and oscillation death in identical coupled oscillators},
  author={Zou, Wei and Senthilkumar, DV and Duan, Jinqiao and Kurths, J{\"u}rgen},
  journal={Physical Review E},
  volume={90},
  number={3},
  pages={032906},
  year={2014},
  publisher={APS}
}

@article{kemeth2019cluster,
  title={Cluster singularity: The unfolding of clustering behavior in globally coupled Stuart-Landau oscillators},
  author={Kemeth, Felix P and Haugland, Sindre W and Krischer, Katharina},
  journal={Chaos: An Interdisciplinary Journal of Nonlinear Science},
  volume={29},
  number={2},
  year={2019},
  publisher={AIP Publishing}
}

@article{kemeth20212,
  title={2-cluster fixed-point analysis of mean-coupled Stuart--Landau oscillators in the center manifold},
  author={Kemeth, Felix P and Fiedler, Bernold and Haugland, Sindre W and Krischer, Katharina},
  journal={Journal of Physics: Complexity},
  volume={2},
  number={2},
  pages={025005},
  year={2021},
  publisher={IOP Publishing}
}

@article{daido2004aging,
  title={Aging transition and universal scaling in oscillator networks},
  author={Daido, Hiroaki and Nakanishi, Kenji},
  journal={Physical review letters},
  volume={93},
  number={10},
  pages={104101},
  year={2004},
  publisher={APS}
}

@article{harrison2020synchronization,
  title={Synchronization in optically trapped polariton Stuart-Landau networks},
  author={Harrison, Stella L and Sigurdsson, Helgi and Lagoudakis, Pavlos G},
  journal={Physical Review B},
  volume={101},
  number={15},
  pages={155402},
  year={2020},
  publisher={APS}
}

@article{gambuzza2020experimental,
  title={Experimental observations of chimera states in locally and non-locally coupled Stuart-Landau oscillator circuits},
  author={Gambuzza, LV and Minati, L and Frasca, M},
  journal={Chaos, Solitons \& Fractals},
  volume={138},
  pages={109907},
  year={2020},
  publisher={Elsevier}
}

@article{klar2023scale,
  title={Scale-free dynamics in the core-periphery topography and task alignment decline from conscious to unconscious states},
  author={Klar, Philipp and {\c{C}}atal, Yasir and Langner, Robert and Huang, Zirui and Northoff, Georg},
  journal={Communications Biology},
  volume={6},
  number={1},
  pages={499},
  year={2023},
  publisher={Nature Publishing Group UK London}
}

@article{hatakeyama2015reciprocity,
  title={Reciprocity between robustness of period and plasticity of phase in biological clocks},
  author={Hatakeyama, Tetsuhiro S and Kaneko, Kunihiko},
  journal={Physical Review Letters},
  volume={115},
  number={21},
  pages={218101},
  year={2015},
  publisher={APS}
}

@article{deco2017dynamics,
  title={The dynamics of resting fluctuations in the brain: metastability and its dynamical cortical core},
  author={Deco, Gustavo and Kringelbach, Morten L and Jirsa, Viktor K and Ritter, Petra},
  journal={Scientific reports},
  volume={7},
  number={1},
  pages={3095},
  year={2017},
  publisher={Nature Publishing Group UK London}
}

@article{deco2015rethinking,
  title={Rethinking segregation and integration: contributions of whole-brain modelling},
  author={Deco, Gustavo and Tononi, Giulio and Boly, Melanie and Kringelbach, Morten L},
  journal={Nature reviews neuroscience},
  volume={16},
  number={7},
  pages={430--439},
  year={2015},
  publisher={Nature Publishing Group UK London}
}

@article{ku2015dynamical,
  title={Dynamical transitions in large systems of mean field-coupled Landau-Stuart oscillators: Extensive chaos and cluster states},
  author={Ku, Wai Lim and Girvan, Michelle and Ott, Edward},
  journal={Chaos: An Interdisciplinary Journal of Nonlinear Science},
  volume={25},
  number={12},
  year={2015},
  publisher={AIP Publishing}
}

@article{pandey2026complex,
  title={Complex dynamics and route to quasiperiodic synchronization in non-isochronous directed Stuart--Landau triads},
  author={Pandey, Ankan and Saha, Sandip and Ghosh, Dibakar},
  journal={Chaos, Solitons \& Fractals},
  volume={204},
  pages={117800},
  year={2026},
  publisher={Elsevier}
}

@article{premalatha2018stable,
  title={Stable amplitude chimera states in a network of locally coupled Stuart-Landau oscillators},
  author={Premalatha, K and Chandrasekar, VK and Senthilvelan, M and Lakshmanan, M},
  journal={Chaos: An Interdisciplinary Journal of Nonlinear Science},
  volume={28},
  number={3},
  year={2018},
  publisher={AIP Publishing}
}

@article{hohlein2019lyapunov,
  title={Lyapunov spectra and collective modes of chimera states in globally coupled Stuart-Landau oscillators},
  author={H{\"o}hlein, Kevin and Kemeth, Felix P and Krischer, Katharina},
  journal={Physical Review E},
  volume={100},
  number={2},
  pages={022217},
  year={2019},
  publisher={APS}
}

@article{ryu2026nonreciprocity,
  title={Nonreciprocity induced spatiotemporal chaos: Reactive vs dissipative routes},
  author={Ryu, Jung-Wan},
  journal={Chaos, Solitons \& Fractals},
  volume={203},
  pages={117647},
  year={2026},
  publisher={Elsevier}
}

@article{shing2013phase,
  title={Phase and amplitude dynamics in large systems of coupled oscillators: Growth heterogeneity, nonlinear frequency shifts, and cluster states},
  author={Shing Lee, Wai and Ott, Edward and Antonsen, Thomas M},
  journal={Chaos: An Interdisciplinary Journal of Nonlinear Science},
  volume={23},
  number={3},
  year={2013},
  publisher={AIP Publishing}
}

@article{shiino1989synchronization,
  title={Synchronization of infinitely many coupled limit-cycle type oscillators},
  author={Shiino, Mastoshi and Frankowicz, Marek},
  journal={Physics Letters A},
  volume={136},
  number={3},
  pages={103--108},
  year={1989},
  publisher={Elsevier}
}

@article{matthews1990phase,
  title={Phase diagram for the collective behavior of limit-cycle oscillators},
  author={Matthews, Paul C and Strogatz, Steven H},
  journal={Physical review letters},
  volume={65},
  number={14},
  pages={1701},
  year={1990},
  publisher={APS}
}

@article{millan2025synchronization,
  title={Synchronization of coupled Stuart-Landau oscillators: How heterogeneity can facilitate synchronization},
  author={Mill{\'a}n, Ana P and Poyato, David and Reynolds, David N and Tudisco, Francesco},
  journal={arXiv preprint arXiv:2510.05243},
  year={2025}
}

@article{benettin1980lyapunov,
  title={Lyapunov characteristic exponents for smooth dynamical systems and for Hamiltonian systems; a method for computing all of them. Part 1: Theory},
  author={Benettin, Giancarlo and Galgani, Luigi and Giorgilli, Antonio and Strelcyn, Jean-Marie},
  journal={Meccanica},
  volume={15},
  number={1},
  pages={9--20},
  year={1980},
  publisher={Springer}
}

@article{sandri1996numerical,
  title={Numerical calculation of Lyapunov exponents},
  author={Sandri, Marco},
  journal={Mathematica Journal},
  volume={6},
  number={3},
  pages={78--84},
  year={1996},
  publisher={Redwood City, Ca.: Advanced Book Program, Addison-Wesley Pub. Co., c1990-}
}

@article{mirollo1990amplitude,
  title={Amplitude death in an array of limit-cycle oscillators},
  author={Mirollo, Renato E and Strogatz, Steven H},
  journal={Journal of Statistical Physics},
  volume={60},
  number={1},
  pages={245--262},
  year={1990},
  publisher={Springer}
}

@article{wolfrum2011spectral,
  title={Spectral properties of chimera states},
  author={Wolfrum, Matthias and Omel’chenko, Oleh E and Yanchuk, Serhiy and Maistrenko, Yuri L},
  journal={Chaos: An Interdisciplinary Journal of Nonlinear Science},
  volume={21},
  number={1},
  year={2011},
  publisher={AIP Publishing}
}

@article{nakagawa1994collective,
  title={From collective oscillations to collective chaos in a globally coupled oscillator system},
  author={Nakagawa, Naoko and Kuramoto, Yoshiki},
  journal={Physica D: Nonlinear Phenomena},
  volume={75},
  number={1-3},
  pages={74--80},
  year={1994},
  publisher={Elsevier}
}

@article{lee2022nontrivial,
  title={Nontrivial twisted states in nonlocally coupled Stuart-Landau oscillators},
  author={Lee, Seungjae and Krischer, Katharina},
  journal={Physical Review E},
  volume={106},
  number={4},
  pages={044210},
  year={2022},
  publisher={APS}
}

@article{premalatha2016imperfectly,
  title={Imperfectly synchronized states and chimera states in two interacting populations of nonlocally coupled Stuart-Landau oscillators},
  author={Premalatha, K and Chandrasekar, VK and Senthilvelan, M and Lakshmanan, M},
  journal={Physical Review E},
  volume={94},
  number={1},
  pages={012311},
  year={2016},
  publisher={APS}
}

@article{zhao2025nonreciprocal,
  title={Nonreciprocal amplification toward chaos in a chain of Duffing oscillators},
  author={Zhao, Luekai and Li, Bojun and Uchida, Nariya},
  journal={Physical Review E},
  volume={112},
  number={1},
  pages={014216},
  year={2025},
  publisher={APS}
}



\end{document}